\pdfoutput=1
\documentclass[12pt,preprint]{aastex}

\usepackage{graphicx}
\usepackage{epstopdf}

\newcommand{\et}{et al.}
\newcommand{\kms}{km s$^{-1}$}
\newcommand{\ha}{H$\alpha$}
\newcommand{\solar}{\ifmmode_{\sun}\else$_{\sun}$\fi}

\newcommand{\HII}{H$\,${\sc ii}}
\newcommand{\HI}{H$\,${\sc i}}

\newcommand{\coldens}{atoms cm$^{-2}$}
\newcommand{\sigdens}{M\solar\ pc$^{-2}$}
\newcommand{\sigcrit}{$\Sigma_{\rm crit}$}
\newcommand{\siggas}{$\Sigma_{\rm gas}$}
\newcommand{\sighi}{$\Sigma_{HI}$}

\newcommand{\sigha}{$\Sigma_{H\alpha}$}

\newcommand{\oh}{$12+\log {\rm (O/H)}$}
\newcommand{\dirr}{dIrr}

\begin{document}

\title{Star Formation in Two Luminous Spiral Galaxies}

\author{Deidre A.\ Hunter\altaffilmark{1}, Bruce G.\ Elmegreen\altaffilmark{2},
Vera C.\ Rubin\altaffilmark{3}, Allison Ashburn\altaffilmark{1,4}, Teresa Wright\altaffilmark{1,5},
Gyula I.\ G.\ J\'{o}zsa\altaffilmark{6,7}, and Christian Struve\altaffilmark{6}}

\altaffiltext{1}{Lowell Observatory, 1400 West Mars Hill Road, Flagstaff, Arizona 86001 USA}
\altaffiltext{2}{IBM T. J. Watson Research Center, PO Box 218, Yorktown Heights, New York 10598 USA}
\altaffiltext{3}{Carnegie Institution of Washington, 5241 Broad Branch Road NW, Washington, DC 20015}
\altaffiltext{4}{Current address: Benedictine College, 1020 North 2nd St., Atchison, KS 66002 USA}
\altaffiltext{5}{Current address: Department of Astronomy, Indiana University, 727 East 3rd St., Bloomington, IN 47405-7105 USA}
\altaffiltext{6}{ASTRON (Netherlands Organisation for Scientific Research NWO),  Oude Hoogeveensedijk 4, 7991 PD Dwingeloo, The Netherlands}
\altaffiltext{7}{Argelander-Institut f\"ur Astronomie, Auf dem H\"ugel 71, D-53121 Bonn, Germany}

\begin{abstract}
We have examined star formation in two very luminous ($M_V=-22$ to $-23$)
Sc-type spiral galaxies, NGC 801 and UGC 2885, using ultra-deep \ha\ images.
We combine these with $UBV$ and 2MASS $JHK$ images and \HI\ maps to explore the star formation characteristics
of disk galaxies at high luminosity.
\ha\ traces star formation in these galaxies to 4-6 disk scale lengths,
but the lack of detection of \ha\ further out is likely due to loss of Lyman continuum photons.
Considering gravitational instabilities alone, we find that the gas and stars in the outer regions are marginally
stable in an average sense, but considering dissipative gas and radial and azimuthal forcing, the outer regions are 
marginally unstable to form spiral arms. 
Star formation is taking place in spiral arms, which are regions of locally higher gas densities.
Furthermore, we have traced smooth exponential stellar disks over 3-orders of magnitude and 4-6 disk scale lengths,
in spite of a highly variable gravitational instability parameter. Thus, gravitational instability thresholds
do not seem relevant to the stellar disk.
One possibility for creating an exponential disk is that
the molecular cloud densities and star formation rates have exponential profiles and this forces the stellar disk to build up such a profile.
Another possibility is that the stellar disk is continuously adjusted to an exponential shape regardless of the star formation profile,
for example through global dynamical process that scatter stars.
However, such scattering processes are only known to operate in spiral systems,
in which case they cannot explain the same dilemma of smooth exponential
disks observed in dwarf irregular galaxies.
\end{abstract}


\section{Introduction} \label{sec-intro}

In the Milky Way we see that stars form from clouds of molecular ${\rm H}_2$, and
there is a correlation between the locations of molecular clouds and newly minted
young stars. Presumably the molecular clouds form from atomic gas, but what are the
conditions for clouds to form in a particular region? Star formation in the inner
regions of spiral galaxies is usually attributed to gravitational instabilities
following Toomre's (1964) condition for collapse (Quirk 1972; Kennicutt 1989).
However, this condition does not explain star formation in the outer parts or in
dwarf irregular (\dirr) galaxies where the gas appears to be more stable. Furthermore, empirical
relationships between gas and star formation (Kennicutt \et\ 2007; Bigiel \et\
2008) become less predictive in dwarfs and the outer parts of spirals where the
range in star formations rates (SFRs) is large for a given total gas density (Bigiel \et\
2010; Barnes \et\ 2012; Ficut-Vicas, in prep).

Yet, stellar disks are traced to extraordinarily large radii, well out into the
accompanying gas disks (see, for example, Ferguson \et\ 1998, Ferguson \& Johnson 2001, Bland-Hawthorn
\et\ 2005, Carraro \et\ 2010, Vlaji\'{c} \et\ 2011, Eufrasio \et\ 2013), and young stars {\it are} found in these outer
disks.  Stars younger than $<$0.5 Gyr are traced to 8 disk scale lengths in NGC 2403
(Barker \et\ 2012) and 10 inner disk scale lengths in M33 (Grossi \et\ 2011). In
NGC 7793 (Radburn-Smith \et\ 2012), M33 (Grossi \et\ 2011), M83 (Bigiel \et\
2010), CIG96 (Espada \et\ 2011), several other spiral galaxies (Barnes \et\ 2012)
and several \dirr\ (Hunter \et\ 2011), star formation continues in the
outer parts down to an \HI\ column density of a few $M_\odot$ pc$^{-2}$ or less. {\it
GALEX} has revealed UV-bright stellar complexes in M83 to 7 kpc (Thilker \et\ 2005).
Even in dwarf galaxies, like the LMC, young stars ($<$few Gyr old) are found to 8
disk scale lengths (Saha \et\ 2010).

Here we discuss star formation in two of the most luminous and massive Sc-type
galaxies in the nearby universe (Burstein \et\ 1982). These galaxies, NGC 801 and
UGC 2885, were originally observed as part of a study of the rotation curves of
spirals by Rubin \et\ (1980). We obtained extremely deep \ha\ images of these
galaxies in order to trace \HII\ regions into the outer disk. These data were
originally taken for the purpose of extending the optical rotation curves further out into the disk,
but here we use them as tracers of star formation.
Characteristics of the galaxies are given in Table \ref{tab-gal}.
We combine the \ha\
data with $UBV$ and Two Micron All Sky Survey (2MASS) $JHK$ images to trace the
stellar populations and \HI\ maps that reveal the atomic gas. With these data we
examine the star formation properties of the galaxies: How far out can we trace
star-forming regions? How does the star formation activity compare with the density
of gas? Are these very luminous galaxies different compared to less luminous
spirals, and how do their outer disks compare with low luminosity dwarf galaxies?
And, what do the data tell us about the disk assembly of luminous spirals?

\section{Observations and Data} \label{sec-obs}

\subsection{\ha\ Imaging}

Deep \ha\ images were obtained 2007 November 6-12 using the T2KB 2048$\times$2048
direct imaging CCD on the 2.1 m telescope at Kitt Peak National Observatory
(KPNO\footnote{Kitt Peak National Observatory, National Optical Astronomy
Observatory, which is operated by the Association of Universities for Research in
Astronomy (AURA) under cooperative agreement with the National Science
Foundation.}). For the on-band filter, we used KPNO \#1495, which has a passband
center at 6691 \AA\ and FWHM of 78 \AA. We used filter \#1564 for the off-band
(stellar continuum). The off-band filter is centered at 6618 \AA\ with a FWHM of 74
\AA. The pixel scale was 0.304\arcsec\ per pixel for a field of view of 10.4\arcmin.
We used observations of the flat-field screen in the dome to correct for
pixel-to-pixel variations. We also observed spectrophotometric standard stars at
several airmasses to correct for extinction by the earth's atmosphere and to
transform counts in the \ha\ images to ergs s$^{-1}$ cm$^{-2}$. The FWHM of stars on
the \ha\ images are 1.5\arcsec\ for NGC 801 and 1.3\arcsec\ for UGC 2885.

For the galaxies, the observing sequence was off-band, \ha, \ha, off-band, \ha, \ha,
off-band, and so forth. Each exposure was 1800 s. There were 24 \ha\ images of NGC
801 and 28 of UGC 2885 for a total on-band exposure time of 12 and 14 hours,
respectively. All images were shifted and scaled to match one of the \ha\ images for
each galaxy. The off-band images were also shifted and scaled to the on-band image
in order to subtract the stellar continuum. The final \ha\ image of NGC 801 has a
stellar FWHM of 1.4\arcsec, and that of UGC 2885 has 1.3\arcsec. The \ha\ image was
geometrically transformed to match the scale and orientation of the $V$-band image
so that we could use the same parameters for surface photometry as was used for the
broad-band imaging.

The rms in the sky-subtracted, edited \ha\ images is 3-4$\times10^{-18}$ ergs s$^{-1}$ cm$^{-2}$.
Thus, the rms includes the sky-subtraction and residuals due to editing foreground stars.
This rms translates to about $3\times10^{36}$ ergs s$^{-1}$ per pixel at the distance of these galaxies.
The FWHM of stars in the \ha\ images is about 2 pixels, and the area covered is about 3 pixels.
Thus, an unresolved \HII\ region at a level of one rms would have a luminosity of about $9\times10^{36}$ ergs s$^{-1}$,
approximately one Orion nebula (Kennicutt 1984).

\subsection{$UBV$ Imaging}

$UBV$ images of NGC 801 and UGC 2885 were obtained 2010 December 1 and 27-28 with an
e2v 2048$\times$2048 CCD on the Lowell 1.1 m Hall telescope. The CCD was binned
2$\times$2 for a pixel scale of 0.74\arcsec. Exposure times were 600 s in $V$, 1200
s in $B$, and 1800 s in $U$. We obtained 10/6/4 exposures in $V/B/U$ of NGC 801 for
total exposure times of 1.7/2.0/2.0 hrs, respectively. For UGC 2885 we have 10/7/6
exposures in $V/B/U$ for total exposure times of 1.7/2.3/3.0 hrs, respectively. We
used observations of the twilight sky to remove pixel-to-pixel variations across the
CCD. We also observed Landolt (1992) standard stars for photometric calibrations.
The rms in the $U$-band calibration is 0.042 mag, for $B$ it is 0.036 mag, and for
$V$ it is 0.028 mag. Stars on the final $V$-band image of NGC 801 have a FWHM of
1.9\arcsec\ and on the UGC 2885 image have a FWHM of 2.4\arcsec.

All of the images in a given filter were aligned, scaled, and averaged to produce a
single final image. The $B$-band image was also shifted to match the $V$-band image
and the $U$-band was geometrically transformed to match $V$. Foreground stars and
background objects were edited from the images and the non-galaxy portion of the CCD
images were fit to determine the two-dimensional background using {\sc imsurfit} in
the {\it Image Reduction and Analysis Facility} (IRAF\footnote{IRAF is distributed
by the National Optical Astronomy Observatory, which is operated by the Association
of Universities for Research in Astronomy (AURA) under cooperative agreement with
the National Science Foundation.}). The background was subtracted from the image to
produce a sky-subtracted image ready for surface photometry. The center of the
galaxy, the position angle (PA) of its major axis, and its minor-to-major axis ratio
$b/a$ were measured from an outer isophote in the $V$-band image. The center of the
galaxy was determined from a radial profile of the nucleus/bulge.

\subsection{$JHK$ Imaging}

We obtained $JHK$ images of the galaxies from the 2MASS online archives of the
All-Sky Release Survey Atlas.  Before performing surface
photometry, we geometrically transformed them to match the scale and orientation of
the $V$-band image, and, as for the other images, edited non-galaxy objects from the
image, fit the 2D background, and subtracted it from the images. The FWHM of stars
in the frames were 2.4\arcsec-2.5\arcsec\ on the NGC 801 images and
2.7\arcsec-2.8\arcsec\ on the UGC 2885 images. The 2MASS images come with
calibration information in their headers.
These images are not very deep, but they give us the
$JHK$ colors of the central regions of the galaxies.

We also examined the {\it Spitzer} 3.6 $\mu$m images of these galaxies, but the
field of view was too small to allow adequate sky-subtraction. Hence, the photometry
was unreliable.

\subsection{\HI-line Data}

\HI-line data were available for both galaxies from the archives of the Westerbork
Synthesis Radio Telescope (WSRT). NGC 801 was observed as part of the Westerbork
Observations of Neutral Hydrogen in Irregular and Spiral Galaxies (WHISP, see van der Hulst \et\ 2001) project
and the data were provided to us. The beam size is 16.1\arcsec$\times$27.0\arcsec,
and the channel separation and spectral resolution is 8.25 \kms. The map cubes are
512$\times$512 pixels and each pixel is 5.0\arcsec.
The rms noise in a channel map of the data cube is $0.73\,{\rm  mJy}\,{\rm beam}^{-1}$.
From the data cube we determine a total flux of $20.96\,{\rm Jy}\,{\rm km}\,{\rm s}^{-1}$,
corresponding to an \HI\ mass of $3.0\times 10^{10}$ M\solar. This is similar to the
\HI\ mass of $2.6 \times 10^{10}$ M\solar, corrected for distance, given by the WHISP project
(http://www.astro.rug.nl/\~{}whisp/).

UGC 2885 was observed with the WSRT in 2004 in two runs with a total on-source integration time of 24 h.
The total bandwidth was 10 MHz with a number of 1024 channels, centered on the \HI\ line of UGC 2885.
We used two parallel-handed polarization products without online Hanning smoothing.
The data underwent a standard data reduction with the Miriad software package (Sault \et\ 1995).
The data were flagged, and a primary bandpass calibration was applied. After that, an iterative self-calibration
on the continuum image was performed to correct the frequency-independent gains. The data were
Hanning-smoothed, continuum-subtracted, gridded and inverted using a set of different weighting schemes,
and subsequently deconvolved employing the CLEAN algorithm. For the latter we used an iterative approach
in which we successively decreased the cutoff level and increased ``clean regions,''
restricting CLEAN to certain areas in the single planes of the cubes. For this study, a data cube employing a
Robust weighting of 0.0 without averaging in frequency was found to be most suitable. The resulting
spatial resolution is 22.3\arcsec $\times$ 13.6\arcsec\ (HPBW), the channel width is
$2.06\,{\rm km}\,{\rm s}^{-1}$, and the rms noise is $0.63\,{\rm mJy}\,{\rm beam}^{-1}$.
The adopted data cube has a size of 256$\times$256 pixels and a pixel size of $4^{\prime\prime}$.
From this data cube, we determined the total \HI\ flux of UGC 2885 to be $28.3\,{\rm Jy}\,{\rm km}\,{\rm s}^{-1}$,
which corresponds to an \HI\ mass of $4.2\times 10^{10}$ M\solar.

\begin{figure}
\epsscale{1.0}
\includegraphics[angle=0,width=1.0\textwidth]{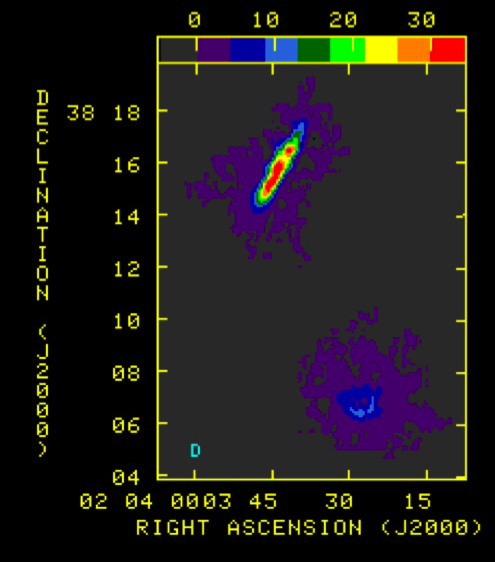}
\caption{False color image of the integrated \HI\ map of NGC 801.
NGC 801 is to the northeast and NGC 797 is the galaxy to the southwest.
The two galaxies are 110 \kms\ and 215 kpc apart at a distance of 79.4 Mpc.
The beam FWHM is shown in the lower left.
The color bar is in units of WU, which is 5 mJy beam$^{-1}$, and 1 WU here corresponds to $1.05\times10^{20}$ \coldens.
\label{fig-n801m0full}}
\end{figure}

Note that there is another galaxy in the field of view of the NGC 801 map. The
galaxy, NGC 797, to the southwest of NGC 801 has a recessional velocity that is 110
\kms\ smaller than NGC 801's and the two galaxies are 215 kpc apart on the sky at a
distance of 79.4 Mpc.  NGC 797 is classed as an SAB(s)a by de Vaucouleurs \et\
(1991). The integrated \HI\ map showing both galaxies is displayed in Figure
\ref{fig-n801m0full}. From here on, only the region immediately around NGC 801 will
be shown.

\section{Analysis} \label{sec-anal}

\subsection{Surface Photometry}

\begin{figure}
\epsscale{1.0}
\includegraphics[angle=0,width=1.0\textwidth]{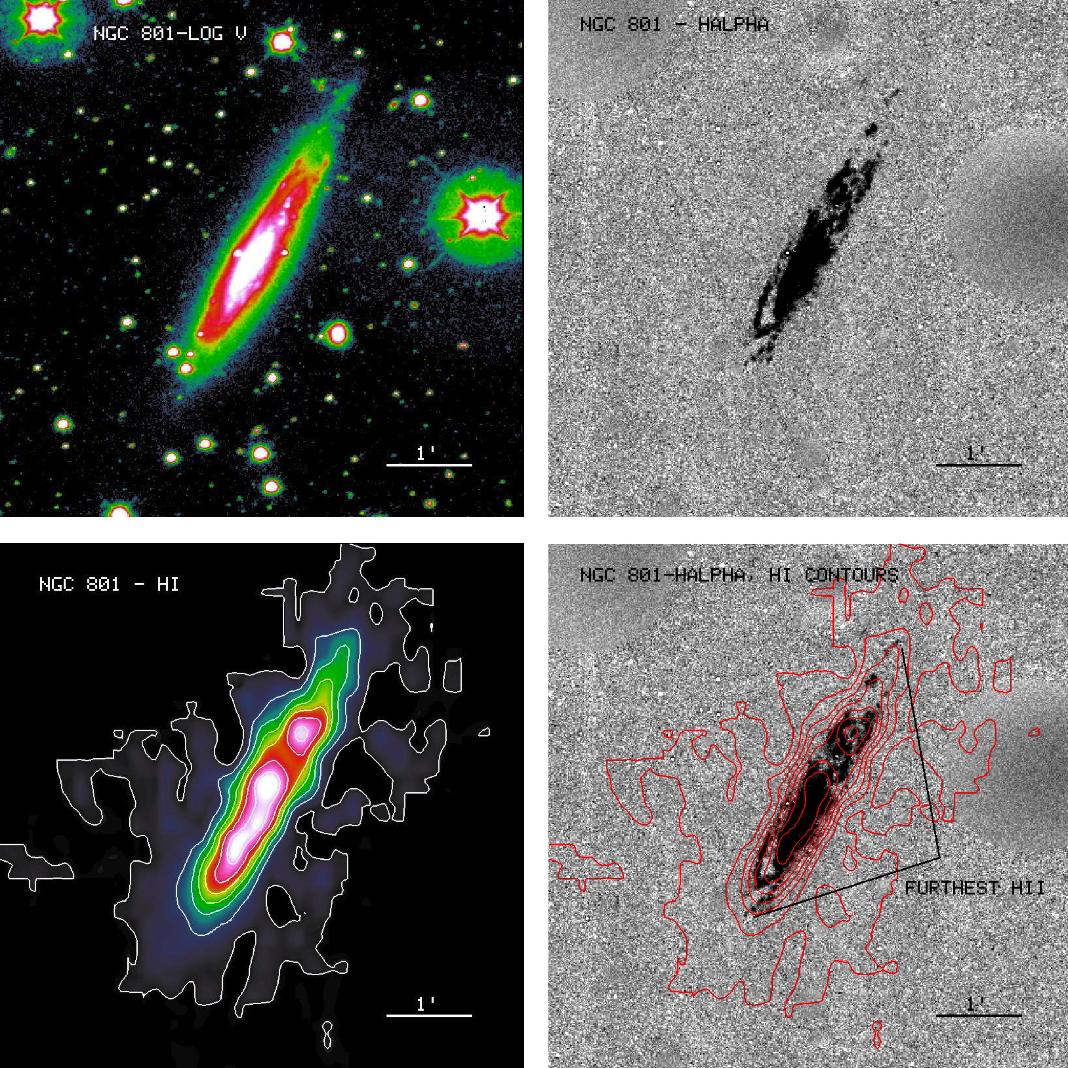}
\caption{Images of NGC 801. All images show the same field of view.
{\it Upper left}: $V$-band shown in false-color as the logarithm in order to allow inner details to be seen.
{\it Upper right}: \ha\ image with stellar continuum subtracted to leave only nebular emission.
{\it Lower left}: Integrated \HI\ map in false-color with contours superposed.
The beam size is 16.1\arcsec\ $\times$ 27.0\arcsec.
{\it Lower right}: \ha\ image with integrated \HI\ contours superposed.
The contours begin at  $4.2\times10^{19}$ cm$^{-2}$ and increase in steps of $5.2\times10^{20}$ cm$^{-2}$.
The location of distant \HII\ regions are marked.
\label{fig-n801color}}
\end{figure}

\begin{figure}
\epsscale{1.0}
\includegraphics[angle=0,width=1.0\textwidth]{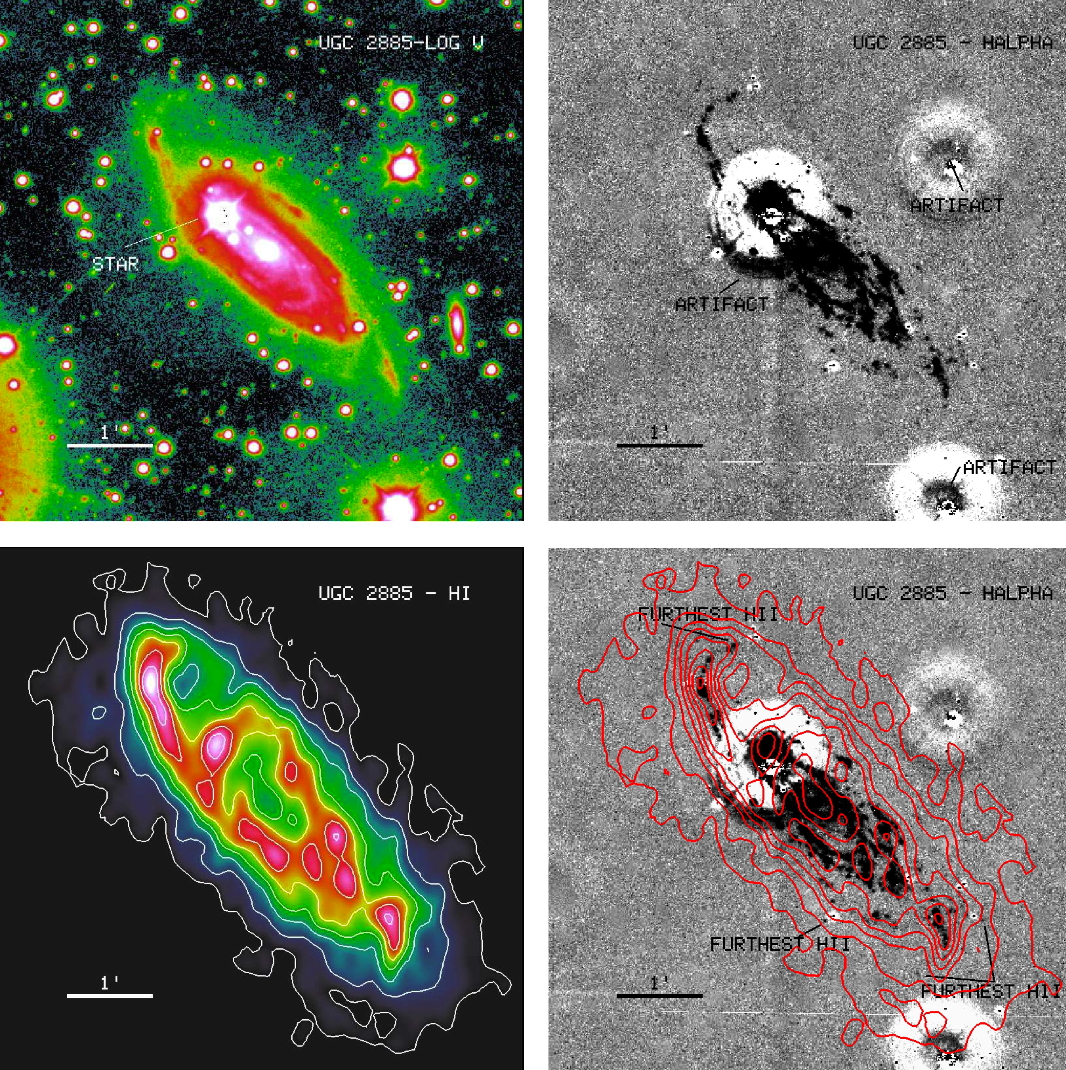}
\caption{Images of UGC 2885. All images show the same field of view.
{\it Upper left}: $V$-band shown in false-color as the logarithm in order to allow inner details to be seen.
{\it Upper right}: \ha\ image with stellar continuum subtracted to leave only nebular emission.
Artifacts due to poor subtraction of saturated stars are marked.
{\it Lower left}: Integrated \HI\ map in false-color with contours superposed.
The beam size is 22.27\arcsec\ $\times$ 13.58\arcsec.
{\it Lower right}: \ha\ image with  integrated \HI\ contours superposed.
The contours begin at  $4.2\times10^{19}$ cm$^{-2}$ and increase in steps of $5.2\times10^{20}$ cm$^{-2}$,
as for NGC 801.
The location of distant \HII\ regions are marked.
\label{fig-u2885color}}
\end{figure}

The $V$-band, \ha, and integrated \HI\ images of NGC 801 and UGC 2885 are shown in
Figures \ref{fig-n801color} and \ref{fig-u2885color}. Both galaxies are seen fairly edge-on;
inclinations are 74\degr\ and 82\degr.

We performed azimuthally-averaged surface photometry on the $UBVJHK$ and \ha\ images
using {\sc ellipse} in IRAF. We began with the edited, sky-subtracted images, all
geometrically transformed to match the scale and orientation of the $V$-band image,
and used the parameters derived from the $V$-band image to define the ellipse shape
and orientation (see Table \ref{tab-gal}). We held the center, PA of the major axis,
and ellipticity of successive ellipses constant and used a step size of 14.8\arcsec\
for the semi-major axis. Annular surface photometry was determined by subtracting
two adjacent ellipses and dividing by the area in the annulus.

Although we edited the images of foreground and background objects, there were some
artifacts from nearby bright stars or fainter stars within the galaxy image that
were masked. UGC 2885, in particular, has a bright star to the northeast of the
galaxy center, so much of that half of the galaxy had to be masked. Masked regions
were replaced with average photometry in the annulus. The azimuthally-averaged
surface photometry, not corrected for reddening, is shown in Figures
\ref{fig-n801nored} and \ref{fig-u2885nored}.

Similar to the stellar photometry, we determined the \HI\ surface densities from the
integrated \HI\ maps using the Groningen Image Processing System (GIPSY, Vogelaar \&
Terlouw 2001). The ellipse parameters and step size were the same as those for the
optical and near-IR images.

\subsection{Extinction and Reddening}

Correcting the surface photometry for extinction due to dust includes foreground
reddening due to the Milky Way E(B$-$V)$_f$ and reddening internal to the galaxy
E(B$-$V)$_i$. The total reddening E(B$-$V)$_t$ is the sum of these two components.
For E(B$-$V)$_f$ we use the values given in the NASA/IPAC Extragalactic Database
(NED) for the recalibration of the Schlegel \et\ (1998) values by Schlafly \&
Finkbeiner (2011). The E(B$-$V)$_f$ are given in Table \ref{tab-gal}. To correct for
the foreground extinction we use the Cardelli \et\ (1989) extinction law that was
determined for the Milky Way: $A_{V,f}=3.1\times$E(B$-$V)$_f$,
E(U$-$B)$_f=0.69\times$E(B$-$V)$_f$, E(J$-$H)$_f=0.27\times$E(B$-$V)$_f$,
E(H$-$K)$_f=0.23\times$E(B$-$V)$_f$, and E(V$-$J)$_f=2.14\times$E(B$-$V)$_f$.

We have no direct measurement of E(B$-$V)$_i$ for our galaxies, and so we estimate
them from observations of other galaxies in the literature. One complication is that
extinction may be a function of galaxy luminosity (Wang \& Heckman 1996) and in
spirals it changes with radius, decreasing into the outer disk (Huizinga \& van
Albada 1992). We began with the work of Prescott \et\ (2007), a study of
star-forming regions in galaxies observed as part of the {\it Spitzer} Infrared
Nearby Galaxies Survey (SINGS). Fitting their Figure 12, we have that the internal
$V$-band extinction of star-forming nebulae is $\log A_{V,i}^{\rm HII} = 0.45 -
0.67\times(R/R_{25})$. Their data only extend to an $R/R_{25}$ of 1.4, so we
extrapolate this to $R/R_{25}$ of 3 for our data. The values from Prescott \et\
(2007) were determined for star-forming regions, and are appropriate for our \ha\
photometry of \HII\ regions. We adopt $A_{H\alpha, i}^{\rm HII}=0.82 A_{V,i}^{\rm
HII}$, which Prescott \et\ took from Calzetti \et\ (2000).

\begin{figure}
\epsscale{1.0}
\includegraphics[angle=0,width=1.0\textwidth]{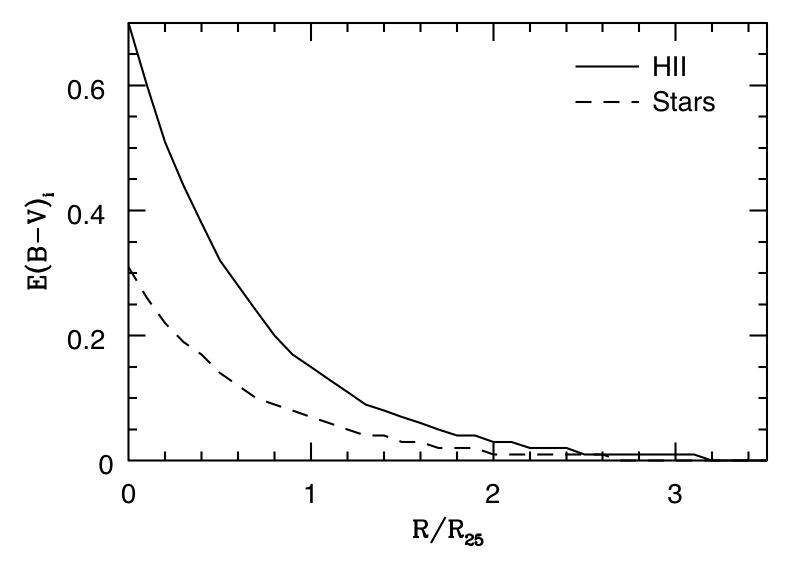}
\caption{Internal reddening E(B$-$V)$_i$ as a function of radius, adopted and extrapolated from Prescott \et\ (2007).
This is appropriate for \HII\ regions, and hence our \ha\ observations.
For the field stars we adopt a scaling factor of 0.44 times E(B$-$V)$_i$
for the \HII\ regions (Calzetti 1994).
\label{fig-ebmv}}
\end{figure}

\begin{figure}
\epsscale{1.0}
\includegraphics[angle=0,width=1.0\textwidth]{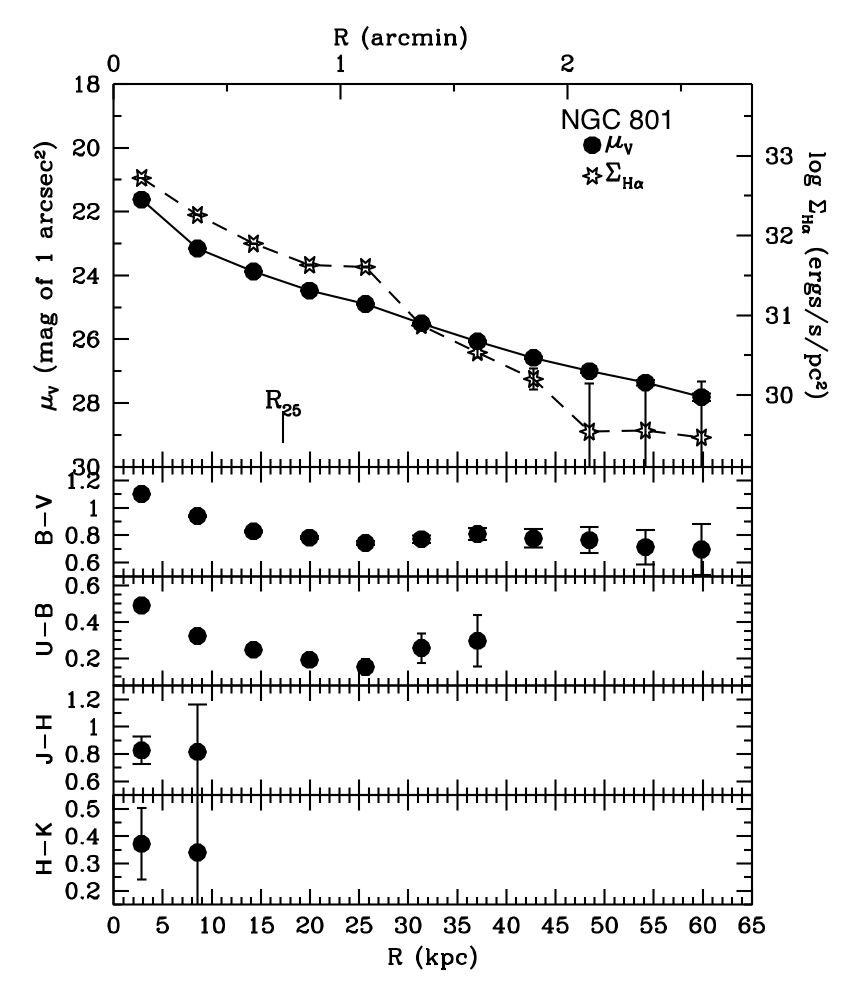}
\caption{Surface photometry and colors of NGC 801 from $UBVJHK$ and \ha\ images.
The photometry has not been corrected for reddening.
$R_{25}$ is the radius of the galaxy at the $B$-band surface brightness of 25 mag of 1 arcsec$^2$.
\label{fig-n801nored}}
\end{figure}

\begin{figure}
\epsscale{1.0}
\includegraphics[angle=0,width=1.0\textwidth]{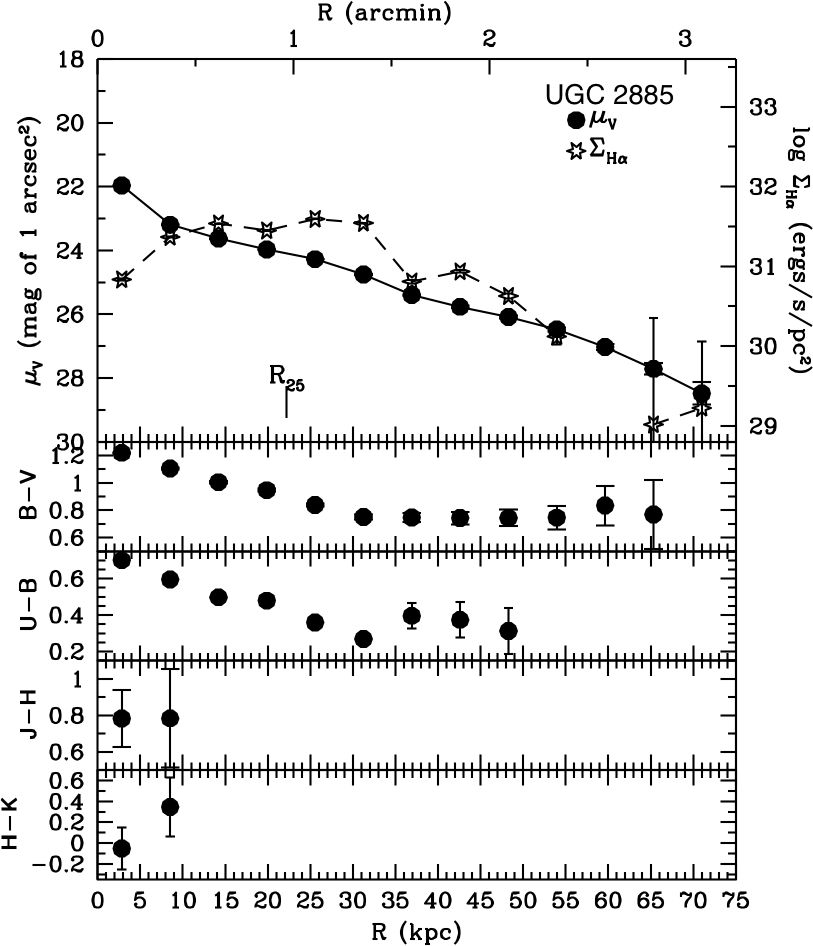}
\caption{Surface photometry and colors of UGC 2885 from $UBVJHK$ and \ha\ images.
There is no emission in the third to last annulus (60 kpc radius) in the azimuthally-average \ha\ photometry;
hence, there is a discontinuity in the \ha\ profile.
The photometry has not been corrected for reddening.
$R_{25}$ is the radius of the galaxy at the $B$-band surface brightness of 25 mag of 1 arcsec$^2$.
\label{fig-u2885nored}}
\end{figure}

We expect field stars to suffer from less extinction than star-forming regions in
dense clouds, and so for the field stars we turn to the extinction law of Calzetti
\et\ (1994, 2000). This extinction law was determined specifically for the stellar
continuum integrated over large regions in starburst galaxies. Our spirals are not
strictly speaking ``starburst'' systems, but they are actively forming stars. For
scaling the extinction law, E(B$-$V)$_i^*$, we use the observation that the
reddening of the stars is related to that of the ionized gas in star-forming
regions, on average, as E(B$-$V)$_i^* = 0.44 \pm 0.03$ E(B$-$V)$_i^{\rm HII}$
(Calzetti 1997). Thus, we use the $A_{V,i}^{\rm HII}$ determined from Prescott \et\
(2007) for the ionized gas, divide that by $R_V=4.05$ (Calzetti \et\ 2000) to obtain
E(B$-$V)$_i^{\rm HII}$, and multiply that by 0.44 to obtain E(B$-$V)$_i^{*}$. The
Calzetti \et\ extinction law can then be used to correct the photometry of the
stellar populations in the various broad-band filters:
$A_{V,i}^*=4.05\times$E(B$-$V)$_i^*$, E(U$-$B)$_i^*=0.95\times$E(B$-$V)$_i^*$,
E(J$-$H)$_i^*=0.57\times$E(B$-$V)$_i^*$, E(H$-$K)$_i^*=0.43\times$E(B$-$V)$_i^*$,
and E(V$-$J)$_i^*=0.97\times$E(B$-$V)$_i^*$. E(B$-$V)$_i^{\rm HII}$ and
E(B$-$V)$_i^*$ are shown as a function of radius in Figure \ref{fig-ebmv}. From this
we have that E(B$-$V)$_i^{\rm HII}$ is $\sim$0.7 mag in the centers of the galaxies
and drops to 0.01 at $R/R_{25}=3$.

For integrated photometry, we use the E(B$-$V)$_i^*$ and E(B$-$V)$_i^{\rm HII}$
appropriate to the $V$-band half-light radius $R_{1/2}$ of the galaxy. $R_{1/2}$ was
determined before any reddening correction was applied. Those values are given in
Table \ref{tab-gal}.

\begin{figure}
\epsscale{1.0}
\includegraphics[angle=0,width=0.7\textwidth]{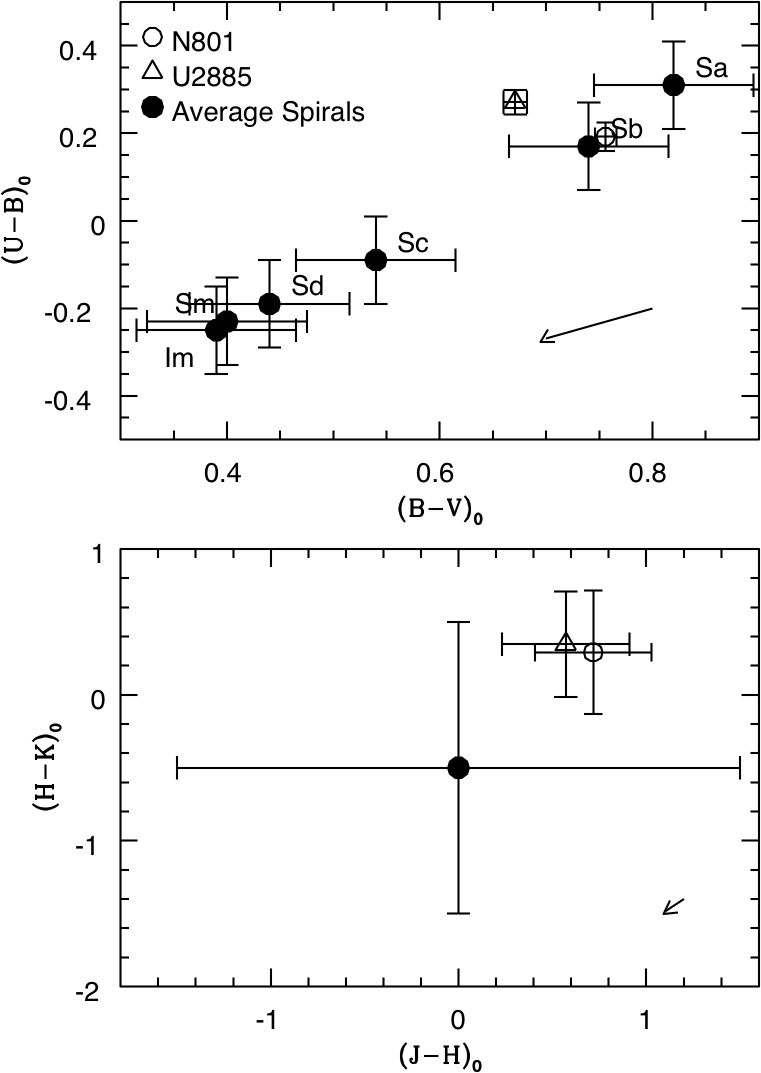}
\caption{Comparison of NGC 801 and UGC 2885 integrated colors to averages for spirals.
{\it Upper panel}: $UBV$ color-color diagram.
Averages by spiral morphological type are from de Vaucouleurs \& de Vaucouleurs (1972).
The uncertainties are the spread in colors for each type.
The arrow shows the reddening vector for E(B$-$V)$=$0.1.
UGC 2885 is redder than other Sc-type galaxies, and NGC 801 has colors comparable to typical Sb-type systems.
{\it Lower panel}: $JHK$ color-color diagram.
The large black point is the average for spiral disks from Holwerda \et\ (2005), and the uncertainties
are the range of values. The arrow is the reddening vector for E(J$-$H)$=$0.1.
NGC 801 and UGC 2885 are redder than most other spirals but within the range of values seen.
\label{fig-comparecolors}}
\end{figure}

Figure \ref{fig-comparecolors} shows the integrated $UBV$ and $JHK$ colors of NGC
801 and UGC 2885 on color-color diagrams. We also show $(U-B)_0$ and $(B-V)_0$
averages by spiral morphological type from de Vaucouleurs \& de Vaucouleurs (1972),
and the uncertainties are the spread in colors for each type. The arrow shows the
reddening vector for E(B$-$V)$=$0.1. NGC 801 and UGC 2885 are redder than other
Sc-type galaxies, comparable instead to Sb-type galaxies. For $(H-K)_0$ vs
$(J-H)_0$, we show the average for spiral disks from Holwerda \et\ (2005) and the
uncertainties are the range of values in the average colors. The arrow is the
reddening vector for E(J$-$H)$=0.1$. NGC 801 and UGC 2885 are redder than most other
spirals in these colors too, but within the range of values seen.

\begin{figure}
\epsscale{1.0}
\includegraphics[angle=0,width=0.8\textwidth]{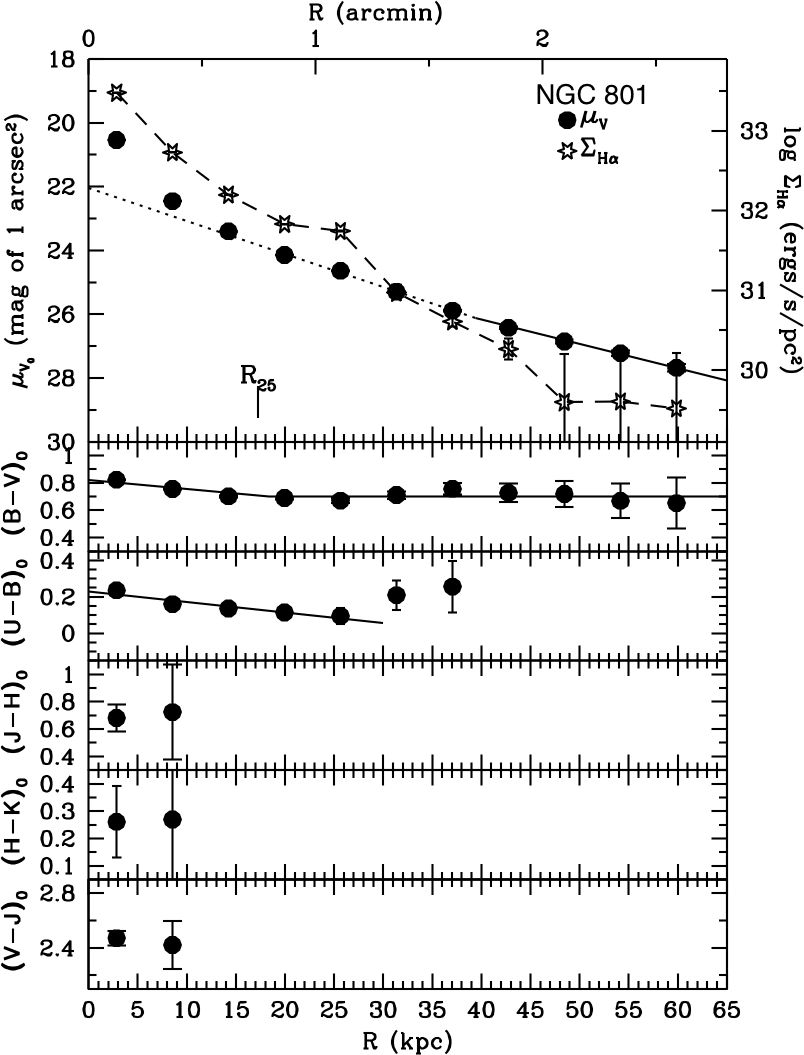}
\caption{Azimuthally-averaged surface photometry and colors of NGC 801 from $UBVJHK$ and \ha\ images,
corrected for reddening.
$R_{25}$ is the radius of the galaxy at the $B$-band surface brightness of 25 mag of 1 arcsec$^2$.
The dotted black line in the upper panel is the fit to the inner $V$-band surface brightness profile,
and the solid line is the fit to the outer profile.
The inner component has a disk scale length of $10.5\pm0.4$ kpc and a central surface brightness
of $22.0\pm0.1$ mag of 1 arcsec$^2$.
The outer component has a disk scale length of $14.1\pm0.5$ kpc and a central surface brightness of the disk
of $23.1\pm0.1$ mag of 1 arcsec$^2$.
The solid lines in the $UBV$ color panels represent a fit to the data: the first 5 points
[(B$-$V)$_0=0.82\pm0.02 - (0.0065\pm0.001)\times{\rm R (kpc)}$, (U$-$B)$_0=0.23\pm0.02 - (0.0058\pm0.001)\times{\rm R (kpc)}$],
and in the case of (B$-$V)$_0$ points 5-11 (constant at 0.70$\pm$0.01).
\label{fig-n801sb}}
\end{figure}

The azimuthally-averaged surface photometry, corrected for reddening, are shown in
Figures \ref{fig-n801sb} and \ref{fig-u2885sb}. In the top panel, $\log
\Sigma_{H\alpha}$ and $\mu_V$ are plotted as a function of radius. The logarithmic
interval is equal to the magnitude interval that is plotted, so that the shapes of
the profiles of the two quantities $\log \Sigma_{H\alpha}$ and $\mu_V$ can be
compared directly.

For NGC 801, the $V$-band surface photometry $\mu_V$ is high in the center,
presumably due to the central bulge. Beyond 20 kpc, $\mu_V$ can be fit with a two-component
exponential with an upbending outer part.
The inner component fits $R=20-40$ kpc and the outer $R=40-60$ kpc, shown in 
Figure \ref{fig-n801sb} as a dotted line and a solid line, respectively.
The inner component is fit with a disk scale length $R_D$ of $10.5\pm0.4$ kpc and a central
surface brightness $\mu_V^0$ of $22.0\pm0.1$ mag arcsec$^{-1}$.
A fit to the outer disk yields a disk scale length $R_D$ of 14.1$\pm$0.5 kpc and a central
surface brightness of $23.1\pm0.1$ mag arcsec$^{-1}$. 
The two components cross at a break radius of 39.4 kpc and a surface brightness of 26.1 mag arcsec$^{-1}$.
Because we emphasize the outer disk here, we use the disk scale length of the outer surface brightness
component.

We see that
$\Sigma_{H\alpha}$ drops like $\mu_V$ with radius until a radius of 26 kpc, and then
it declines faster. \ha\ ends where the $V$ surface photometry ends. The $V$-band
image is not particularly deep, so it is possible that $\mu_V$ continues to lower
surface brightness levels, but the \ha\ exposure {\it is} deep and it is unlikely
that \ha\ would be detected much further out. In NGC 801 the $(B-V)_0$ and $(U-B)_0$
colors are red in the center and get slightly ($\sim$0.15 mag) bluer to a radius of about 25 kpc,
approximately the same radius where the \ha\ surface brightness begins to decline
faster than $\mu_V$. Beyond that radius $(B-V)_0$ stays constant within the
uncertainties, and $(U-B)_0$ gets slightly ($\sim$0.16 mag) redder. There are only
two annuli with $JHK$ colors, but they are red and within the uncertainties constant over
that limited radius range.

For UGC 2885, there is a small excess of $\mu_V$ in the inner 8.5 kpc, but beyond
that we see an exponential disk. The disk scale length is 12.0$\pm$0.4 kpc and
$\mu_V^0$ is 21.3$\pm$0.1 mag in $V$. Here we see that $\Sigma_{H\alpha}$ is low in
the center, and then from 8.5 kpc onward it declines with radius like $\mu_V$. \ha\
emission is traced to 74 kpc, similar to $\mu_V$, which is traced to 71 kpc. The
colors  $(B-V)_0$ and $(U-B)_0$ are red in the center and get bluer ($\sim$0.25, 0.23 mag) with radius to 31
kpc. Beyond that radius they stay approximately constant within the uncertainties,
although the $(U-B)_0$ color exhibits a 0.1 mag offset to the red. There are only
two annuli with $JHK$ colors, but the colors are red in the center, and in $(H-K)_0$ and $(V-J)_0$ the colors
are a bit redder in the second annulus compared to the first.

\subsection{Determining Star Formation Rates}

We use \ha\ to determine the SFRs in NGC 801 and UGC 2885 using the formula of
Kennicutt (1998). However, first we must correct the flux measured in the \ha\
filter for the contribution by [NII]$\lambda$6548+6583 emission. We must also
correct the SFR formula for the metallicity of the stars since the output of
ionizing photons changes with metallicity for a given stellar mass.

\subsubsection{[NII] Correction}

The \ha\ filter bandpass includes [NII]$\lambda$6548 and [NII]$\lambda$6584, and we
need to subtract the contribution from [NII] to the flux in order to derive pure
\ha\ emission. First, all three emission lines are located on the flat part of the
transmission curve of the filter. That means that there is no difference in the
transmission of the different emission lines, and we need only correct for the
intrinsic relative fluxes of the emission lines. Second, from the \ha\ survey of
James \et\ (2005), 5 spirals give a global average ratio \ha/(\ha$+$[NII]) of 0.82,
where [NII] is the sum of the emission from $\lambda$6548 and $\lambda$6583.
So, we multiply our \ha\ fluxes by 0.82 to correct for [NII] emission.

\subsubsection{Metallicity}

Spiral galaxies typically exhibit abundance gradients with radius. For NGC 801 and
UGC 2885 we adopt M81's oxygen abundance gradient as given by Patterson \et\ (2012):
$-0.117\pm0.073$ dex R$_{25}^{-1}$.
M81 has an integrated $M_V$ of $-20.8$, so it is less luminous than NGC 801 and UGC
2885 by a factor of 3-6. However, Henry \& Worthey (1999) argue that abundance
gradients in spirals are independent of the absolute magnitude of the galaxy when the gradient is
expressed in terms of radius normalized to $R_{25}$. We use 
Patterson's ``KK04'' gradient, determined from 4 data sets, which has the advantage that
the oxygen abundance \oh\ at the center is 9.0. This is close to
the maximum \oh\ of 8.9 found by Pilyugin \et\ (2007) at the
centers of luminous ($M_B= -22.5$) spirals.
Therefore, we adopted \oh $= 9.0 - 0.12(R/R_{25})$ as the oxygen abundance as a function of radius in
our spirals.
This is shown in Figure \ref{fig-metal}. \oh\ starts at 9.0 in the
center and drops to 8.58, 78\% solar, at $R/R_{25}=3.5$. We use a solar oxygen
abundance of 8.69 (Asplund \et\ 2009).

\subsubsection{The SFR Formula}

To convert $\Sigma_{H\alpha}$ in units of ergs s$^{-1}$ pc$^{-2}$, as shown in
Figures \ref{fig-n801sb} and \ref{fig-u2885sb}, to SFR in units of M\solar\
yr$^{-1}$ kpc$^{-2}$, we begin with the formula of Kennicutt (1998): ${\rm
SFR}_{H\alpha} {\rm (M\solar\ yr^{-1})} = 7.9\times10^{-42} L_{H\alpha} {\rm (ergs ~
s^{-1})}$. We multiply the \ha\ surface density by 0.82 to remove [NII] and then
adjust the formula for the deviation of the metallicity from solar, the abundance
that Kennicutt used. For the latter, we use the synthesis models of STARBURST99
(Leitherer \et\ 1999), with constant SFR and a Salpeter (1955) stellar initial mass
function. Their Figure 78, which gives the number of photons below 912 \AA, enables
us to find the average flux of ionizing photons per second from a stellar population
relative to that at solar and so to modify Kennicutt's \ha\ formula as a function of
$Z$ from twice solar to below solar. 
The correction factor at twice solar metallicity is 1.18 and at 40\% solar is 0.87. 
We interpolate to determine the correction factor for the metallicity at a given radius.
The derived SFRs for NGC 801 and UGC 2885 are shown in Figure \ref{fig-sfr}. 

Integrated SFRs use the oxygen abundance at the
half-light radius to determine the SFR formula. For both galaxies this is approximately
$z=0.03$. Thus, the SFRs need to be multiplied by a factor of 1.089 to adjust
for the metallicity.
The integrated SFRs are given in Table \ref{tab-gal}.

We have also computed a SFR necessary to form the stellar mass that currently exists
in each galaxy, SFR$_V$. For this we used the $V$-band surface brightness to
determine the mass in stars and assume a timescale of 12 Gyr for the formation of
that stellar mass. We obtain $M/L_V$ as a function of (B$-$V)$_0$ from Bell \& de
Jong (2001). If the uncertainty in the color is  greater than 0.1 mag, we use the
average  (B$-$V)$_0$ in the outer disk. SFR$_V$ is the SFR necessary to produce the
original mass in stars. Because the $M/L_V$ yields the mass currently in stars, we
correct this by a factor of two for the mass recycled back into the interstellar
medium (Brinchmann \et\ 2004) in order to use the total mass ever formed into stars
in computing the lifetime average ${\rm SFR}_V$. The ratio ${\rm SFR}_{H\alpha}/{\rm
SFR}_V$ is shown in Figure \ref{fig-rat}. We also compute the current surface
density of the stellar mass $\Sigma_*$ in units of M\solar\ pc$^{-2}$.

\section{Discussion} \label{sec-disc}

\subsection{Extent of \ha\ emission}

The extent of \ha\ emission in our two galaxies was determined by eye on the continuum-subtracted images. 
The furthest emission we detected in NGC 801 to the north of the galaxy has a
surface brightness of $10\times10^{-18}$ ergs cm$^{-2}$ s$^{-1}$ arcsec$^{-2}$.
In UGC 2885 the faintest region we traced to the southwest has an \ha\ surface brightness
of $5\times10^{-18}$ ergs cm$^{-2}$ s$^{-1}$ arcsec$^{-2}$. 
This represents the faintest surface brightness we could probably identify as a distinct
region.

In NGC 801 the \ha\ disk extends further to the northwest than to the southeast, and
the furthest discrete \HII\ region that we can identify is found at a radius of $\sim$55 kpc. The
southeast \HII\ regions only extend to 43 kpc. The local \sighi\ is
$1\times10^{21}$ \coldens\ at the radius of the most distant \HII\ region.  This HI
column density is integrated along the line of sight for a highly inclined disk.
Multiplying by the cosine of the inclination, $82^\circ$, to convert to the
perpendicular surface density, we get $1.4\times10^{20}$ \coldens\ $=1.5\;M_\odot$
pc$^{-2}$ including He and heavy elements. The outer \HII\ regions in NGC 801 fall along the extensions of spiral
arms. That star formation even in the outer disks of spirals is generally found
associated with spiral arms seems to be a general feature of spiral galaxies
(Ferguson \et\ 1998, Thilker \et\ 2007, Barnes \et\ 2012).  Outer star formation in \dirr\ galaxies, by contrast,
is not in spiral arms, but in local clumps of high gas surface density (Hunter 1982, Hunter \& Gallagher 1986).

On the northeast end of the disk of UGC 2885 there is an arc of \HII\ regions,
probably the end of a spiral arm, the tip of which is 68 kpc from the galactic
center. The local \sighi\ there is $2\times10^{21}$ \coldens\, which is
$6.0\;M_\odot$ pc$^{-2}$ after correcting for inclination and heavy elements.
On the southwest side of
the galaxy, there are two possible faint
detached \HII\ regions at 60 kpc and 74 kpc. Off the major axis,
there is a region to the southeast of center located at a de-projected radius of 74 kpc.
These detached regions are located at \sighi\ of
$8\times10^{20}$ \coldens\ ($2.3\;M_\odot$ pc$^{-2}$ face-on).

In Figures \ref{fig-n801himasssfr} and \ref{fig-u2885himasssfr} we show the
azimuthally-averaged \ha-based SFRs, $\Sigma_{\rm SFR, H\alpha}$, along with \sighi\
and the stellar mass surface density $\Sigma_*$. We see that the SFR drops 4 dex in
NGC 801 from the center to a value of $10^{-6}$ M\solar\ yr$^{-1}$ kpc$^{-2}$ at a
radius of 60 kpc. In UGC 2885 it drops 1.5 dex to a value of $10^{-5}$ M\solar\
yr$^{-1}$ kpc$^{-2}$ at 54 kpc. These very low SFRs in the outer disks are
comparable to those seen in less luminous Sab to Sd-type outer disks (Barnes \et\
2012) and in the outer disks (3-5 kpc radius) of dwarf galaxies (Hunter \et\ 2011).

Some of the drop in \ha\ at large radii could be from a loss of Lyman continuum
photons from the vicinity of star formation (Beckman \et\ 2000 [but see Lee \et\ 2011],
Hunter \et\ 2010, Pellegrini \et\ 2012). If we re-write the
Str\"{o}mgren relation as
\begin{equation}
nR_{\rm S}=( 3S/ [ 4 \pi R_{\rm S} \alpha ] )^{1/2}
\end{equation}
for ambient density $n$, Str\"{o}mgren radius $R_{\rm S}$, stellar ionization rate $S$,
and case B recombination rate $\alpha$, then the critical column density for trapping Lyman
continuum radiation from an O-type star with $S=10^{49}$ s$^{-1}$, considering
$\alpha\sim2.6\times10^{-13}$ cm$^3$ s$^{-1}$ (Osterbrock 1989) and $R_{\rm
S}=100R_2$ pc is
\begin{equation}
nR_{\rm S,crit} = 1.7\times10^{20}R_2^{-1/2} \;{\rm cm}^{-2}=1.9
R_2^{-1/2} \;M_\odot\;{\rm pc}^{-2}.
\label{eq:crit}
\end{equation}
To consider the possible loss of Lyman continuum photons from a region of star
formation, the column density through a critical Str\"{o}mgren diameter, $2nR_{\rm
S,crit}$, should be compared to the local average gas column density measured
perpendicular in the disk.
Equation (\ref{eq:crit}) suggests that the critical value for trapping
is $4\;M_\odot$ pc$^{-2}$ or less for a single O-type star, especially if $R_2$ gets
large in a disk flare.
In fact, the most distant \HII\ regions in our
galaxies have local \sighi\ at about this limit or less. This suggests we could be
missing other \HII\ regions at these galactocentric distances and certainly missing
those much further out. As a result, our estimates of star formation rates from
outer disk H$\alpha$ should be considered lower limits.

The loss of ionizing radiation at extremely low \sighi\ is in addition to the loss
studied by Rela\~no \et\ (2012) that is for visible \HII\ regions. They found
that shell \HII\ regions have lower \ha\ compared to FUV than compact regions,
indicating up to a 25\% loss of ionization in the shell phase. If \sighi\ is as low
as $\sim2\;M_\odot$ pc$^{-2}$, however, then even these visible \HII\ regions should
be difficult to see.  First, there should be a loss of radiation from the disk
because the Str\"{o}mgren radius is larger than the scale height, which is the condition
for the critical disk column density derived above.

Second, the emission from the fully
ionized disk itself will be difficult to see because the emission measure is low.
This fully ionized emission measure depends on the gas column density and scale height. The
scale height is $\sigma^2/\pi G \Sigma_{\rm total}$ for velocity dispersion $\sigma$ and total
mass column density in gas and stars, $\Sigma_{\rm total}$. We assume  $\sigma=10$ km s$^{-1}$  and
take the total column densities from Figures \ref{fig-n801himasssfr} and \ref{fig-u2885himasssfr}. For NGC 801
at the outer annulus in the azimuthally-averaged surface photometry, $R\sim60$ kpc,
$\Sigma_{\rm total}=0.6\;M_\odot$ pc$^{-2}$, which gives a scale height of 13.1 kpc.
For UGC 2885 at its
outer annulus for H$\alpha$ detection, $R\sim71$ kpc,  $\Sigma_{\rm total}\sim1.3\;M_\odot$ pc$^{-2}$, and the
scale height is 5.7 kpc.

\begin{figure}
\epsscale{1.0}
\includegraphics[angle=0,width=0.8\textwidth]{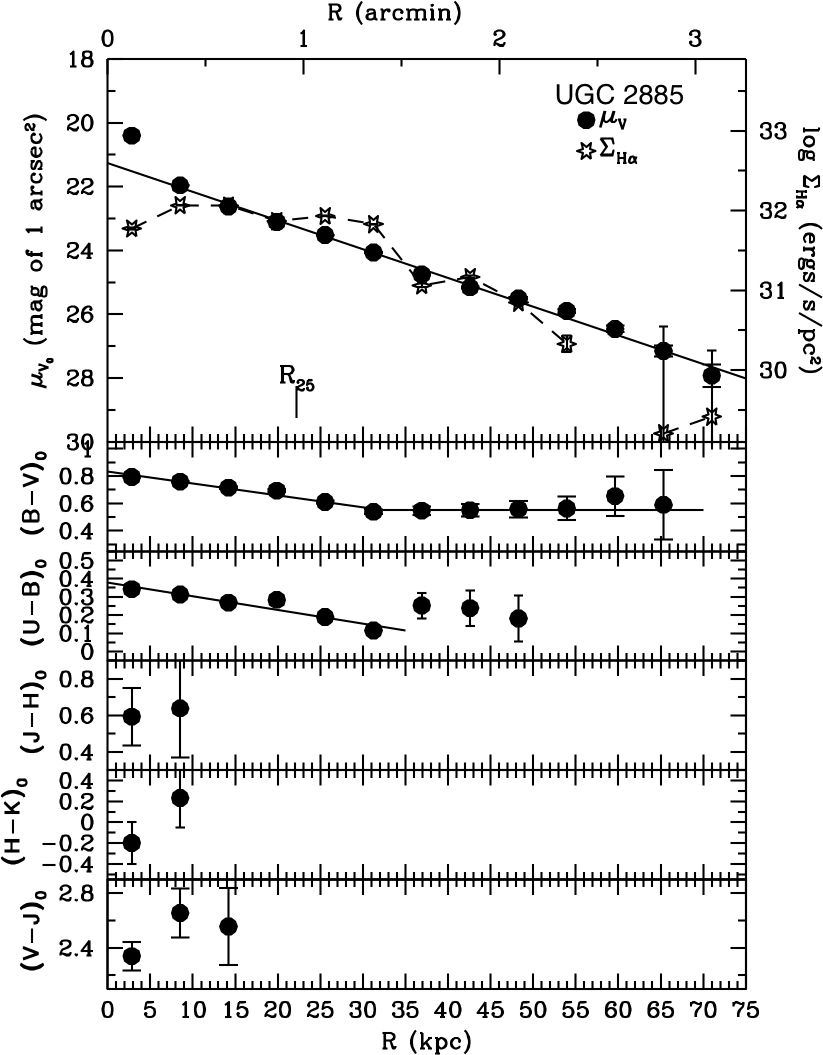}
\caption{Azimuthally-averaged surface photometry and colors of UGC 2885 from $UBVJHK$ and \ha\ images,
corrected for reddening.
There is no emission in the third to last annulus (60 kpc radius) in the azimuthally-average \ha\ photometry;
hence, there is a discontinuity in the \ha\ profile.
$R_{25}$ is the radius of the galaxy at the $B$-band surface brightness of 25 mag of 1 arcsec$^2$.
The solid black line in the upper panel is the fit to the outer $V$-band surface brightness profile.
This yields a disk scale length of 12.05$\pm$0.41 kpc and a central surface brightness of the disk
of 21.26$\pm$0.15 mag of 1 arcsec$^2$.
The solid lines in the $UBV$ color panels represent a fit to the data: the first 6 points
((B$-$V)$_0=0.83\pm0.02 - (0.0087\pm0.001)\times{\rm R (kpc)}$, (U$-$B)$_0=0.38\pm0.03 - (0.0075\pm0.001)\times{\rm R (kpc)}$],
and in the case of (B$-$V)$_0$ points 6-10 (constant at 0.55$\pm$0.004).
\label{fig-u2885sb}}
\end{figure}

\begin{figure}
\epsscale{1.0}
\includegraphics[angle=0,width=1.0\textwidth]{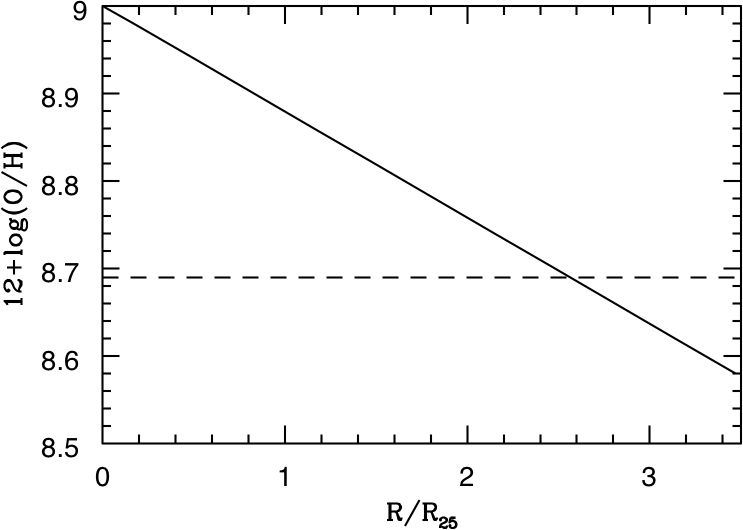}
\caption{Oxygen abundance as a function of radius adopted from Patterson \et\ (2012) for M81
and extrapolated to our radial extent.
The dashed horizontal line marks the solar abundance of 8.69 (Asplund \et\ 2009).
\label{fig-metal}}
\end{figure}

\begin{figure}
\epsscale{1.0}
\includegraphics[angle=0,width=1.0\textwidth]{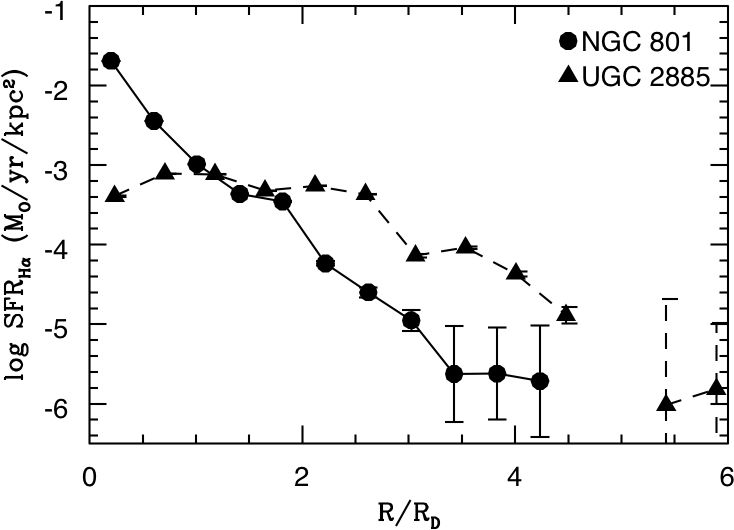}
\caption{H$\alpha$-derived SFRs for NGC 801 and UGC 2885, where the formula for converting
\ha\ surface photometry to SFR$_{H\alpha}$ is modified for the abundance gradient
adopted in Figure \protect\ref{fig-metal}.
For UGC 2885, there is no \ha\ emission in the third to last azimuthally-averaged annulus (60 kpc radius);
hence, there is a discontinuity in the SFR$_{H\alpha}$ profile.
\label{fig-sfr}}
\end{figure}

\begin{figure}
\epsscale{1.0}
\includegraphics[angle=0,width=1.0\textwidth]{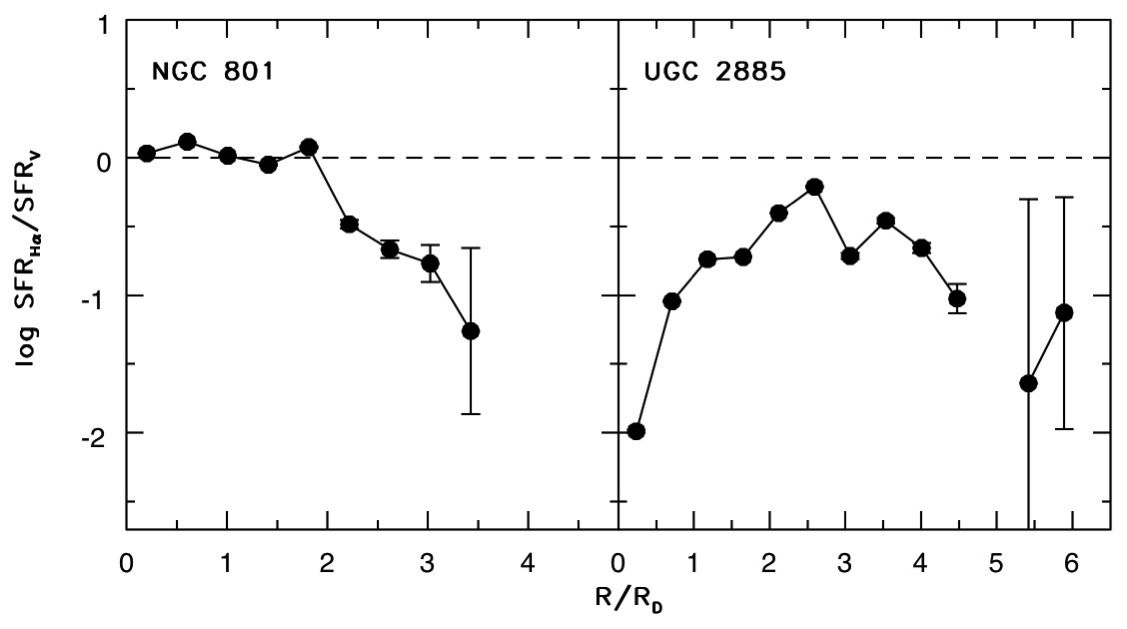}
\caption{SFR derived from H$\alpha$ divided by the lifetime-averaged SFR derived from the stellar mass, determined from the $V$-band
surface brightness corrected to face-on. The horizontal dashed line at 0 marks the value for equal SFRs. The current star formation activity,
SFR$_{H\alpha}$, is everywhere much lower than that averaged over the lifetime of the galaxies, SFR$_V$.
For UGC 2885, there is no \ha\ emission in the third to last azimuthally-averaged annulus (60 kpc radius);
hence, there is a discontinuity in the SFR$_{H\alpha}$ profile.
\label{fig-rat}}
\end{figure}

\begin{figure}
\epsscale{1.0}
\includegraphics[angle=0,width=1.0\textwidth]{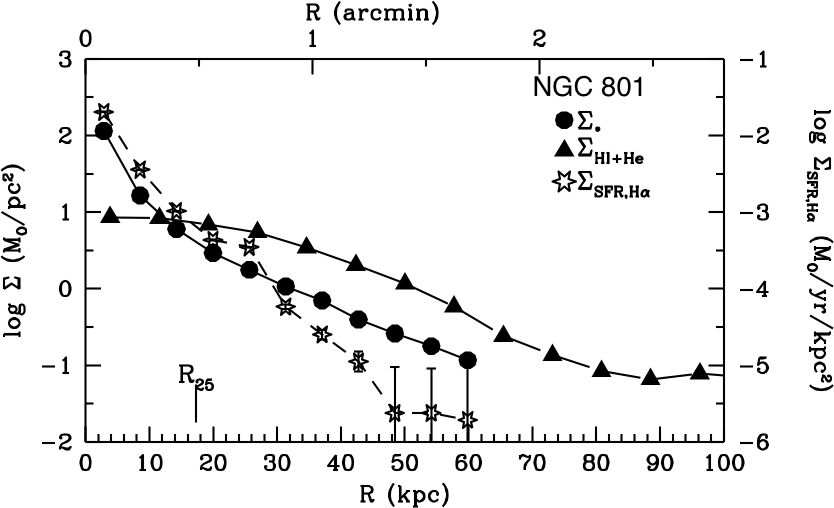}
\caption{Stellar mass surface density $\Sigma_*$,
\HI$+$He surface density $\Sigma_{\rm HI+He}$, and SFR density $\Sigma_{\rm SFR, H\alpha}$ plotted as a function of radius
for NGC 801.
The gas and stellar mass surface densities have been corrected to face-on.
The logarithmic interval is the same for all three quantities, but the SFR zero point is different.
\label{fig-n801himasssfr}}
\end{figure}

\begin{figure}
\epsscale{1.0}
\includegraphics[angle=0,width=1.0\textwidth]{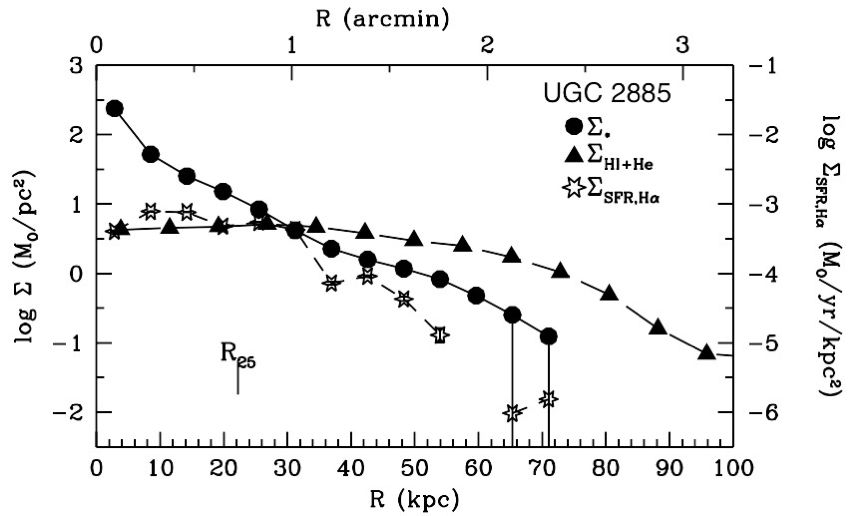}
\caption{Stellar mass surface density $\Sigma_*$,
\HI$+$He surface density $\Sigma_{\rm HI+He}$, and SFR density $\Sigma_{\rm SFR, H\alpha}$ plotted as a function of radius
for UGC 2885.
The gas and stellar mass surface densities have been corrected to face-on.
The logarithmic interval is the same for all three quantities, but the SFR zero point is different.
There is no \ha\ emission in the third to last azimuthally-averaged annulus (60 kpc radius);
hence, there is a discontinuity in the SFR$_{H\alpha}$ profile.
\label{fig-u2885himasssfr}}
\end{figure}

The full disk thicknesses are twice these scale heights. If we take the measured
$\Sigma_{\rm gas}=0.45\;M_\odot$ pc$^{-2}$ at this radius in NGC 801 and correct for
He and heavy elements, then the average \HI\ density over the full thickness is
0.00052 cm$^{-3}$ and the emission measure would be 0.0069 cm$^6$ pc if the disk is
completely ionized. Similarly, for UGC 2885, where $\Sigma_{\rm gas}=1.18\;M_\odot$
pc$^{-2}$,  the density and emission measure would be 0.0031 cm$^{-3}$ and 0.11
cm$^6$ pc. Emission measure in \ha\ converts to \sigha\ as shown in Figures
\ref{fig-n801nored}, \ref{fig-u2885nored}, \ref{fig-n801sb}, and \ref{fig-u2885sb}
according to the equation
\begin{equation}
\Sigma_{H\alpha} \;({\rm erg\;s^{-1}\;pc^{-2}}) = 7.7\times10^{30} EM\;({\rm cm^6\;pc})
\end{equation}
so the full-disk emission measures at the outer radii of the \HII\ regions in these
two galaxies convert to log-luminosity densities of $\log\Sigma_{\rm H\alpha}=28.7$
for NGC 801 and 29.9 for UGC 2885. In fact these are close to the minimum average
luminosity densities for H$\alpha$ at the outer radii of these galaxies (Figures
\ref{fig-n801sb} and \ref{fig-u2885sb}). We conclude again that we are probably
missing \ha\ emission and underestimating the \ha\ SFR near the edges of these
galaxies.

In NGC 801 we see that $\Sigma_{\rm SFR, H\alpha}$ decreases with radius in the
inner region like the surface density of the stellar mass even though the excess
mass in the center is due to the galaxy bulge. Beyond a radius of 25 kpc,
$\Sigma_{\rm SFR, H\alpha}$ drops faster than $\Sigma_*$. This type of behavior
is seen in other spirals as well (Christlein \et\ 2010).
However, beyond 25 kpc the
\HI\ gas surface density is about 4 times the stellar surface density and they both drop at
nearly the same rate as far as our data trace the stars.
Beyond our $V$-band image, the gas continues to decrease gently.

In the center of UGC 2885, there is a dip in the value of $\Sigma_{\rm SFR, H\alpha}$.
Beyond about 15 kpc radius, the azimuthally-averaged $\Sigma_{\rm SFR, H\alpha}$
falls, more or less, like the surface density of the mass in stars, until 48 kpc.
Beyond that radius $\Sigma_{\rm SFR, H\alpha}$ drops more rapidly than $\Sigma_*$ until
both stars and \ha\ emission are no longer detected at a radius of 71 kpc.
At a radius of about 30 kpc, the \HI\ and stellar mass surface densities are equal, and
beyond that the stellar mass surface density drops more steeply than the gas,
which continues to decrease gently, as in NGC 801.
This lack of a correlation between outer disk star formation and outer disk gas is
also evident in galaxies with large \HI\ disks studied by Wang \et\ (2013; in prep).

In NGC 801 and UGC 2885, we trace \ha\ emission as far out as we trace $V$-band starlight.
However, it is possible that young stars
might be detectable in FUV emission even beyond where \ha\ emission
ends, since in outer disks UV light from young massive stars traces recent star
formation more easily than \ha\ emission from nebulae (as discussed above, and see,
for example, Thilker \et\ 2005, 2007; Boissier \et\ 2007). Unfortunately, FUV images
of these galaxies do not exist.

\subsection{Radial trends in SFRs}

In Figure \ref{fig-rat} we show the ratio of current ($\leq10$ Myr) SFR to the
lifetime averaged SFR necessary to form the mass in stars, ${\rm SFR}_{H\alpha}/{\rm
SFR}_V$, over 4-6 disk scale lengths.
We find that the ratio is near 1 in the central regions of NGC 801, to a radius of 1.8$R_D$.
Beyond that in NGC 801 and everywhere in UGC 2885, the current SFR is significantly
lower than the lifetime-averaged rate.
For NGC 801 the ratio drops to 0.09 at a radius of 4$R_D$.
For UGC 2885 it is quite low ($\sim$0.01) in the center, rises to a high value of 0.6
at a radius of 2.6$R_D$, and then drops slowly to a value of 0.07 at a radius of 6$R_D$.
Some of the drop in the perceived SFR in the far-outer regions could be from a loss
of sensitivity to very faint \HII\ regions there, considering that the \HI\ column
density is comparable to or lower than the critical value (Eq.\ \ref{eq:crit}) for
trapping ionization in the nearby ambient medium.

In a deep imaging study of 5 \dirr\ galaxies we found a different situation (Hunter
\et\ 2011). In three of the dwarfs the ratio of current to lifetime-average SFR is
roughly 1 at all radii. In the other two galaxies, the ratio is of order 0.1, so the
current SFR is low compared to the past average. This low ratio is comparable to the
maximum values in NGC 801 and UGC 2885. So, compared to dwarfs, these luminous
spirals have formed much more of their stellar mass in the past than they are adding
today.

The trends in NGC 801 and UGC 2885 are different from the general relationship seen
in some spirals, as well. In a large sample of spirals, Ryder \& Dopita (1994),
found that the disk scale-length in H$\alpha$ is much longer than that of the
$V$-band, and the $V$-band scale length is longer than that of the $I$-band surface
photometry. Similarly, Chang \et\ (2012) found in one gas-rich spiral that the
recent SFR relative to the lifetime average increased as one moves to the outer
disk. These types of observations are interpreted as evidence for an inside-out disk
assembly model (e.g. Larson 1976; Chiappini \et\ 1997; Mo \et\ 1998; Naab \&
Ostriker 2006). 

In UGC 2885 the inner part of the galaxy appears to be dominated by older stars while the
outer disk is relatively younger, consistent in a general sense with inside-out growth. On the other hand,
the current SFR drops relative to the lifetime average beyond a radius of 2.5$R_D$.
So, while the outer disk is younger than the central regions overall, the outer disk becomes more and more
dominated by older stars with increasing radius.
In NGC 801, however, the inner disk exhibits a constant SFR and then
the current SFR drops steadily in the outer disk, beyond a radius of 2$R_D$, compared to
the lifetime average. This implies that the outer disk is relatively older than the
inner disk, counter to inside-out disk growth models. 

Yoachim \et\ (2012) also find mixed star formation profiles in a sample of six nearby spirals:
half are dominated by an older stellar population beyond a break in the surface
brightness profile, which they interpret as evidence for radial migration of older
stars. Other studies have also found evidence for extended old stellar disks (see,
for example, Ferguson \& Johnson 2001, Vlaji\'{c} \et\ 2009). The other half of the
Yoachim \et\ sample shows no significant change in the  stellar population in the
outer disk. Thus, there does not seem to be a universal star formation profile among
spirals (see also, Roediger \et\ 2012).

The decrease in the ratio of current to past star formation in the outer regions of our galaxies, particularly in
NGC 801, contradicts our expectation from the lack of a $B-V$ color gradient, which is shown in Figures
\ref{fig-n801sb} and \ref{fig-u2885sb}. We seem to require some radial migration
of stars to make the color gradients flat  (e.g., Ro\u{s}kar \et\ 2008).

\subsection{The Toomre model}

\begin{figure}
\epsscale{1.0}
\includegraphics[angle=0,width=1.0\textwidth]{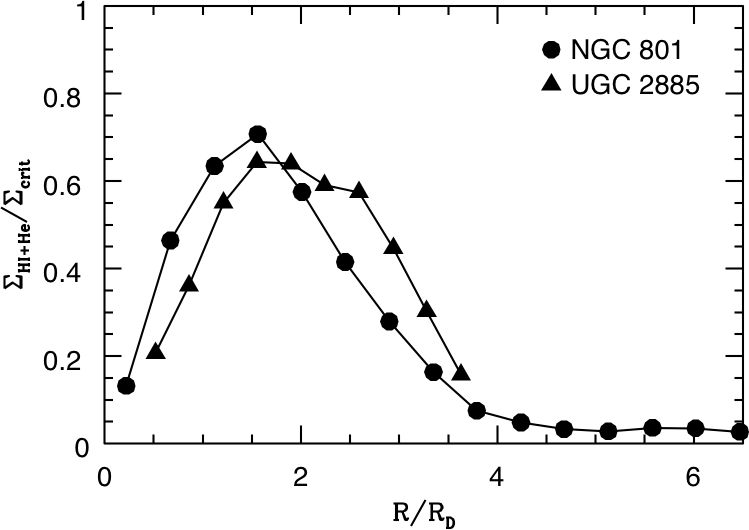}
\caption{Ratio of \HI$+$He gas surface
density to critical gas density predicted by the gravitational
instability model of Toomre (1964) versus radius in units of disk scale length.
\label{fig-sigcrit}}
\end{figure}

Toomre (1964) calculated the dispersion relation of self-gravitating waves in a thin
rotating stellar disk and found a minimum velocity dispersion in the radial
direction for stability. This critical velocity dispersion has found popular use in
star formation models when it is re-written as a critical column density for gas
\sigcrit. If the gas column density is greater than \sigcrit, then the model
suggests star formation is active because of widespread instabilities. To check this
model, we calculated \sigcrit\ as a function of radius for our galaxies using the
rotation curves in Rubin \et\ (1980). \sigcrit\ depends on the rotation curve
through the epicyclic frequency $\kappa$: \sigcrit$ =\kappa\sigma_{\rm g}/\pi G$ for
gas velocity dispersion $\sigma_{\rm g}$. We assume $\sigma_{\rm g}=10$ \kms,
a typical spiral galaxy gaseous velocity dispersion (Tamburro \et\ 2009).
We also assume that the velocity dispersion is independent of radius, although there is 
evidence that the velocity dispersion is lower in
outer spiral disks (Kamphuis \& Sancisi 1993, Tamburro \et\ 2009).
The affect of lowering the velocity dispersion would be to 
decrease \sigcrit, making the gas proportionally more unstable against gravitational instabilities.
The ratios of the observed \siggas\ to \sigcrit\ are shown for our two galaxies in Figure \ref{fig-sigcrit}.

In the conventional instability model (e.g., Kennicutt 1989), stars should form at a
greater rate where $\Sigma_{\rm gas}/\Sigma_{\rm crit}$ is higher. This is
approximately true for our galaxies too. In both cases the ratio of the observed
\siggas\ to \sigcrit\ is low in the center, climbs to a peak at mid-radius, and then
drops in the outer disk.
The radius at which \siggas/\sigcrit\ begins to drop rapidly (1.8-2.5$R_D$)
is also the same radius at which the current SFR drops relative to
the integrated past SFR (Figure \ref{fig-rat}).
However, this pattern is usually found in exponential
disks with flat rotation curves, because then $\Sigma_{\rm gas}/\Sigma_{\rm crit}
\propto R \exp(-R/R_{\rm D})$ for constant velocity dispersion regardless of star
formation, and this function has the same general rising and falling shape as the
observed profile. The low values of $\Sigma_{\rm gas}/\Sigma_{\rm crit}$ in the
centers of our galaxies could be from a high $\kappa$, a lack of gas, or the
presence of molecules that are not included in \siggas.
The maximum value of the
ratio in NGC 801 is 0.7 and in UGC 2885 it is 0.6. At these peaks, the gas is close
to the original Toomre (1964) critical density. However, at the maximum radius where
\HII\ regions are detected, the ratio is 0.2 in both NGC 801 and UGC 2885, which
is far into the conventionally stable regime (see also the spirals in Barnes \et\
2012).

The low values of $\Sigma_{\rm gas}/\Sigma_{\rm crit}$ in the outer regions mean
that stars are forming in gas that is sub-critical in the usual sense. This is
similar to the situation in the outer disks of \dirr\ galaxies, even though the
average \HI\ column densities where the last \HII\ regions are found in our spirals
are higher than those in the outer disks of dwarfs by a factor of 10 (Hunter \et\
2011). Clearly, it is not \HI\ column density alone that determines where stars form
(but see Bigiel \et\ 2010).

We also see no threshold or shift in the star formation rate at some characteristic
$\Sigma_{\rm gas}/\Sigma_{\rm crit}$. The underlying stellar disk has a smooth,
featureless and extended exponential profile, with no features at some radius where
the star formation rate or disk instability may be changing properties. If this
exponential disk reflects the past history of star formation, then there is no
evident threshold.

The stability of galaxy disks is more complicated than what can be captured by this
one-dimensional model, however. Stars contribute to the instability, disk thickness
impedes it, gas dissipation promotes it, and spirals that fragment into
self-gravitating clouds can still form in sub-threshold conditions.

A combined gas+star disk is more unstable than a pure gas disk (Toomre 1964; Jog \&
Solomon 1984). Romeo \& Wiegert (2011) suggest an effective $Q_{\rm eff}$ given by
\begin{equation}
Q_{\rm eff}=\left( {{W}\over{T_{\rm s}Q_{\rm s}}}+ {{1}\over{T_{\rm g}Q_{\rm
g}}}\right)^{-1} \label{eq:qeff}
\end{equation}
in the likely case where $T_{\rm s}Q_{\rm s}>T_{\rm g}Q_{\rm g}$, which means that
the stellar disk alone is more stable than the gaseous disk alone. Here,
$T=0.8+0.7\sigma_{\rm z}/\sigma_{\rm R}$ is a thickness correction for gas or stars
with velocity dispersions $\sigma$ in the perpendicular ($z$) or radial ($R$)
directions. Romeo and Wiegert suggest $\sigma_{\rm z}/\sigma_{\rm R}=0.6$ for stars
and 1 for gas, making $T_{\rm s}=1.22$ and $T_{\rm g}=1.5$. Also, $W=2\sigma_{\rm
s}\sigma_{\rm g}/\left(\sigma_{\rm s}^2+\sigma_{\rm g}^2\right)$.
The quantities $T$ and $W$ depend only on the ratios of velocity dispersions, 
so a decrease with radius in these dispersions will not affect the results of this discussion. 
As for the dispersion in \sigcrit, the values used in $Q_{\rm s}$, $Q_{\rm g}$ and $W$ are radial components.

Note that $Q_{\rm eff}<1$ becomes the instability condition for a 2-fluid thick
disk, replacing $Q_{\rm g}<1$, or equivalently, $\Sigma_{\rm gas}/\Sigma_{\rm
crit}>1$, for the thin disk model discussed above.  Thus, we can convert $Q_{\rm
eff}$ to a threshold 2-fluid gas column density for thick disks, $\Sigma_{\rm
crit,2F,thick}$ by setting $Q_{\rm eff}=1$ in Eq.\ \ref{eq:qeff}. We write the
1-fluid critical value for infinitely thin disks that we have been using as
$\Sigma_{\rm crit,1F,thin}=\kappa \sigma_{\rm g}/(\pi G)$. Then, inverting Eq.\
\ref{eq:qeff} and substituting for $W$, we get
\begin{equation}
\Sigma_{\rm crit,2F,thick}=
\Sigma_{\rm crit,1F,thin}T_{\rm g}- {{2\Sigma_{\rm star}T_{\rm g}}\over
{(1+\sigma_{\rm s}^2 /\sigma_{\rm g}^2)T_{\rm s}}}
\end{equation}
Using the above values for the thickness corrections, $T$, we get a ratio of
observed gas column density to the critical, two-fluid, thick disk column density
\begin{equation}
{{\Sigma_{\rm HI+He}}\over{\Sigma_{\rm crit,2F,thick}}}=
{{{\Sigma_{\rm HI+He}}/{\Sigma_{\rm crit,1F,thin}}}\over {1.5-0.488\Sigma_{\rm s}/
\Sigma_{\rm crit,1F,thin}}},
\end{equation}
where we have assumed $\sigma_{\rm s}/\sigma_{\rm g}=2$.
Figures \ref{fig-n801himasssfr} and \ref{fig-u2885himasssfr} suggest that in the
outer disks of our two spiral galaxies, $\Sigma_{\rm s}\sim\Sigma_{\rm g}/4$, and
Figure \ref{fig-sigcrit} indicates that $\Sigma_{\rm HI+He}/\Sigma_{\rm
crit,1F,thin}\sim0.2$ or less there. Thus, $\Sigma_{\rm s}/ \Sigma_{\rm
crit,1F,thin}\sim0.05$, and for this value, the denominator becomes 1.47. Then
\begin{equation} {{\Sigma_{\rm HI+He}}\over{\Sigma_{\rm crit,2F,thick}}}\sim0.68
{{\Sigma_{\rm HI+He}}\over{\Sigma_{\rm crit,1F,thin}}}.
\label{eq:p68}
\end{equation}
This result means that the 2-fluid thick disk is more stable than the 1-fluid thin
disk considered in Figure \ref{fig-sigcrit}, primarily because of thickness effects,
even with the presence of stars. The outer disk is therefore 30\% more
stable than indicated in Figure \ref{fig-sigcrit}.

Energy dissipation changes this picture, however. Two-component star+gas thick disks
with turbulent gas dissipation (Elmegreen 2011) are more unstable than 2-fluid thick
disks that follow the Toomre model, which for gas, assumes an isothermal fluid. For
typical conditions in galaxies, instabilities that create star-forming clouds in
turbulent gas can occur with average gas column densities as low as 30-50\% of the
Romeo \& Wiegert (2011) critical value given above.  This range assumes that
turbulent energy dissipates on between 1 and 2 crossing times over the length of the
self-gravitational perturbation, and it assumes a ratio of velocity dispersions
$\sigma_{\rm s}/\sigma_{\rm g}$ between 1 and 5. Lower $\sigma_{\rm g}/\sigma_{\rm
s}$ and faster dissipation relative to a crossing time correspond to lower critical
gas column densities in the 2-fluid case; i.e., they correspond to greater
instability. This factor of 2 or 3 increase in the ratio of the observed
$\Sigma_{\rm HI+He}$ to the threshold column density, now with dissipation included,
is still not enough to bring us to an unstable regime in the far outer parts, however.

So far, all of these gravitational instability models apply only to the radial
direction, and the instabilities make only rings spaced by the Jeans length. Before
a disk becomes this unstable because of radial forcing alone, it becomes unstable to
the formation of spiral arms by a combination of radial and azimuthal forcing.
Spiral arm formation should always be viewed as the first step in star formation
when a disk has shear (which spiral galaxies have but \dirr\ galaxies do not).
Clouds should not be expected to form directly from the interstellar medium in
random places, but they should form as condensations in the spiral arms.

The stability condition for combined radial and azimuthal forcing was written in
terms of an effective $Q$ parameter by Lau \& Bertin (1978) and Bertin \et\ (1989;
see also Julian \& Toomre 1966),
\begin{equation}
Q_{\rm eff}=\left({{1}\over{Q^2}}+{{J^2Q^4}\over{4}}\right)^{-1/2}
\label{q2}
\end{equation}
where $J$ is a parameter such that
\begin{equation}
{{J^2Q^4}\over 4}=\left(q_{\rm t}Q\right)^2
\left({{2\Omega}\over{\kappa}}\right)^2 \left(\mid{{
d\ln\Omega}\over{d\ln R}}\mid\right)\sim2\left(q_{\rm t}Q\right)^2
\end{equation}
for rotation rate $\Omega$ and dimensionless azimuthal wavenumber
$q_t=2\pi\sigma/(\lambda_{\rm t} \kappa)$ at wavelength $\lambda_{\rm t}$. The
factor 2 in the last expression is for a flat rotation curve.  If there are $m$
spiral arms in the azimuthal direction, then $q_{\rm t}=\sigma m / (\kappa R)$, and
if we write $Q=\sigma \kappa / (\pi G \Sigma)$ in the usual way, then $q_{\rm
t}Q=\sigma^2 m/(\pi G \Sigma R)\sim Hm/R$ for scale height $H$. Condition (\ref{q2})
was applied to disk stability and star formation by Zasov \& Simakov (1988).
For $\sigma\sim10$ km s$^{-1}$ and $\Sigma$ from Figures \ref{fig-n801himasssfr}
and \ref{fig-u2885himasssfr}, we showed above
that $H=13.1$ kpc at $R=60$ kpc for NGC 801 and $H=5.7$ kpc at 71 kpc for UGC 2885,
respectively, in which case ${{J^2Q^4}/{4}}\sim8(H/R)^2\sim0.38$ and $0.05$ for
$m=2$.  If we now use $\Sigma_{\rm HI+He}/\Sigma_{\rm crit,1F,thin}\sim0.13$ and 0.34 from
Figure \ref{fig-sigcrit} at these outer radii for the two galaxies, and take
$2\times0.68$ of this ratio for $\Sigma_{\rm HI+He}/\Sigma_{\rm crit,2F,thick}$ from
equation (\ref{eq:p68}), considering a factor of 2 for turbulence dissipation, then
$Q$, which is the inverse of this ratio, becomes $\sim5.7$ for NGC 801 and 2.1 for
UGC 2885, which are well in the stable regime. However, the azimuthal force lowers
these values considerably, to $Q_{\rm eff}\sim1.6$ and 1.9, respectively. Perhaps as a result of
this azimuthal forcing, spiral arms dominate the morphology of star formation even
in the outer regions.

The above discussion suggests that for realistic assumptions about the stellar disk
thickness and dissipative gas, and with combined radial and azimuthal forcing, the
outer regions of our galaxies are marginally unstable to form spirals. Star
formation inside of these spirals is greatly aided by the higher local densities.
Similarly, Herbert-Fort \et\ (2012) suggest that young ($\leq 1$ Gyr) clusters
formed near the outer edges of galaxy disks are helped by spiral arms. They also
note that star clusters much further out, near the \HI\ edge, may have formed as a
result of gas accretion from the intergalactic medium. It is possible that multiple
factors are responsible for facilitating star formation in the outer disk. In the
case of NGC 801, we also cannot rule out gravitational perturbations by the
neighboring galaxy as a means of generating spiral arms. Also, propagation of spiral
waves from the inner galaxy could be causing the spiral arms in the outer galaxy
without requiring a local mechanism from locally unstable conditions.

\subsection{Star formation thresholds}

With \ha, we have azimuthally-averaged star formation profiles to 4$R_D$ in NGC 801 and 6$R_D$ in UGC 2885,
and we have traced exponential stellar disk profiles in both galaxies to the same radii.
What is striking is the
lack of any disruption in the smooth decline of stellar surface brightness with radius in spite
of the highly variable instability parameter $Q$ (see Figure \ref{fig-sigcrit}).
We argued in the previous section that the traditional $Q$ parameter overestimates the
stability of the gas, but here we point out that $Q$ really does not seem relevant at all
to the stellar disk.
There is no change in the exponential profiles; they drop smoothly with radius
and with no observed end.
Others have also found no correlation between star formation efficiency and $Q$
in inner (Leroy \et\ 2008) or outer (Bigiel \et\ 2010) spiral disks.
Furthermore, the end to \ha\ emission, too, is not likely related to $Q$, but rather to a lack of detectability,
as discussed previously. Earlier inferences of star formation thresholds based on \ha\ emission
did not consider the increased difficulty in detecting \ha\ emission in outer disks
and since then, FUV emission in spirals has shown that there is star formation in
far outer disks, well beyond where \ha\ emission is no longer seen (see, for
example, Thilker \et\ 2007).

The same situation applies to dwarf galaxies as well, even though their gas disks and $Q$ profiles are even more
extreme than in spirals. Our ultra-deep imaging of 5 \dirr\ galaxies
traced the stellar disks to a $V$-band magnitude of 30 mag arcsec$^{-2}$ (Hunter \et\ 2011), and several of these galaxies
show no change in their exponential profile to that surface brightness.
Some dwarfs and spiral galaxies do have breaks in the slopes of
their stellar surface brightness profiles (see, for example, Herrmann \et\ 2013),
but there is no obvious correlation between the radial location of the breaks and a threshold in the $Q$ radial profile.

\subsection{Populating the outer disk and disk formation}

NGC 801 and UGC 2885 have smooth exponential profiles out to galactocentric
distances of at least 60--70 kpc with no apparent cut-off.  The \HII\ regions are
also observed out this far in these galaxies, although the average surface brightness of
\ha\ drops before the optical edge.

The origin of exponential disk profiles is not well understood (van der Kuit \&
Freeman 2011), and the current observations present severe constraints on any model.
The rotation speed in the outer part of NGC 801 is $\sim220$ km s$^{-1}$ (Rubin \et\
1980), so the orbit time at the outer limit of the exponential disk that we see
here, 60 kpc, is 1.7 Gyr. UGC 2885 has an outer disk rotation speed of 300 km
s$^{-1}$ (Rubin \et\ 1980; Roelfsema \& Allen 1985), so at its outer observed
limit, 72 kpc, the orbit time is 1.5 Gyr. These times are $\sim12$\% of a Hubble
time, so the outer disks could have rotated around only $\sim8$ times at their
greatest possible ages. Any process that forms an exponential disk would have to be faster than this.

The same orbit times are derived for the four galaxies in our deep study of
\dirr\ that have outer disk rotation curves (Hunter \et\ 2011), DDO 86: 3.7
Gyr, DDO 133: 1.1 Gyr, NGC 4163: 1.3 Gyr, and IZw 115: 3.0 Gyr. These galaxies also
have nearly perfect exponential disks out to the observed edges, where the surface
brightness is 30 mag arcsec$^{-2}$ in $V$-band -- fainter than the present
observations.

In the outer regions of our spiral galaxies, $\Sigma_{\rm total}\sim1\;M_\odot$
pc$^{-2}$ (Figures \ref{fig-n801himasssfr} and \ref{fig-u2885himasssfr}), so if the
perpendicular velocity dispersion is $\sigma\sim10$ km s$^{-1}$, then the crossing
time through the disk is $\sigma/(\pi G \Sigma_{\rm total})\sim720$ Myr. A complete
cycle perpendicular to the disk involves 4 of these crossings of a scale height
each, so the time is again $\sim3$ Gyr.

The dynamical time for star formation is also about the disk crossing time, i.e.,
several hundred Myr in the far-outer disks studied here. Thus, we have outer disk
star formation that appears analogous to inner disk star formation except for an
extremely low average rate. The average rate decreases with radius by nearly 3
orders of magnitude if we scale it to the surface brightness.
One order of magnitude
of this decrease is because the dynamical time is longer in the outer part, another
order of magnitude is because the total gas surface density is lower in the outer
part, and a third order of magnitude is because the fraction of the gas that is
molecular is lower in the outer part
if the areal star formation rate is proportional to the molecular column density.

It seems baffling how the average star formation rate can maintain such a precise
exponential disk for 3 orders of magnitude in both \dirr\ galaxies and
these spiral galaxies. This is especially puzzling when the molecular fraction is
only a small part of the gas: how do molecular clouds forming at extremely low rates
know what the rest of the galaxy is doing?
The orbit time varies with radius like a
power law ($1/R$) while the perpendicular oscillation time or disk crossing time,
which is approximately the turbulent dissipation time,
increases exponentially with
radius for a fixed velocity dispersion. In the inner disk, the perpendicular
crossing time is much less than the orbit time, but in the outer disk these times
are about the same.  The Toomre model has a characteristic timescale comparable to
the epicyclic time, which is about the orbit time,
whereas cloud formation that
combines gas dissipation and local stellar dynamics has a timescale somewhat
between the orbit and crossing times.
If these times vary separately with radius,
then how do the cloud formation rate and internal cloud evolution rate combine to
give a star formation rate that ends up precisely exponential?

If only the perpendicular crossing time mattered for star formation, because the
most important processes are governed by dissipation, for example, then the
exponential can be self-sustaining: once the disk has an exponential dependence on
radius, then the perpendicular crossing time is exponential, and star formation
following this crossing time would also be exponential. This situation could be
stabilized in the short term by feedback, which puffs up the disk when the star
formation rate is too high and makes the crossing time longer.
But it would seem to
be unstable in the long term for a given average velocity dispersion because a high
star formation rate in some annulus would make the disk relatively dense there, and this
would lower the local crossing time and speed up the star formation, making the disk even
denser.
Subsequent smoothing by stellar spirals (Sellwood \& Binney 2002) could recover the original exponential.
The same instability would seem to arise in more detailed models, such as
that by Ostriker \et\ (2010), in which the star formation rate per unit area
scales with the gas column density multiplied by the square root of the local
midplane stellar volume density.
If the long-term star formation rate increases with
background stellar density, then the exponential profile would seem to be unstable.

We are left with a choice between two puzzling models: (1) the molecular cloud and star
formation rates have an exponential profile independent of the underlying stars and
this rate profile forces the stellar disk to build up such a profile too, or, (2) the
stellar disk is continuously adjusted to an exponential shape regardless of the star
formation profile, and then star formation tends to follow the existing stars for
dynamical reasons and occasional sequential triggering.
However, the latter seems more
reasonable at the present time because global dynamical processes can be imagined
which scatter the stars and keep the disk globally exponential.  Stellar waves might
do this (Roskar \et\ 2008), although dwarf irregulars have extended exponentials
with no obvious waves. Initial conditions that are exponential (Mestel 1963, van der
Kruit 1987) do not help solve this problem because in the present models of galaxy
formation, disk mass is added by gas accretion over a long period of time with in
situ star formation playing a decisive role in the distribution of new stars.
In both of these models, the star formation profile does not follow the gas in the outer parts,
which is mostly \HI.

\section{Conclusions} \label{sec-conclusions}

We have examined star formation in two very luminous ($M_V=-22$ to $-23$)
Sc-type spiral galaxies, NGC 801 and UGC 2885, using ultra-deep \ha\ images.
We combine these with $UBV$ and 2MASS $JHK$ images and \HI\ maps to explore the star formation characteristics
of disk galaxies at high luminosity.

\ha\ traces star formation in these galaxies to 4-6 disk scale lengths.
We find that the most distant \HII\ regions are found at the tips of spiral arms: at radii of 3.9$R_D$ and 5.6$R_D$
and local \HI\ densities of 1.5 \sigdens\ and 6 \sigdens, for NGC 801 and UGC 2885, respectively.
In UGC 2885 there are also a few identifiable detached \HII\ regions as far out as 6.1$R_D$ at 2.3 \sigdens.

Except in the center ($R<1.8R_D$) of NGC 801, the azimuthally-averaged current SFR is lower than
the SFR necessary to form the stellar mass over the lifetime of the galaxy and the current SFR is particularly low
in the outer disks. However, the decrease in current SFR relative to that in the past is not consistent with the relatively
constant colors in the outer disks.
The lack of detection of \ha\ further out is likely due to loss of Lyman continuum photons,
and stellar migration processes may also be important in populating the outer disk.

The gas in the outer regions is marginally stable to gravitational instabilities in an average sense, and star formation is
taking place in regions of locally higher gas densities, namely, spiral arms.
The traditional Toomre (1964) gravitational instability parameter $Q$ is everywhere larger than 1, implying a stable gas.
However, when we include the
stellar disk thickness with dissipative gas and both radial and azimuthal forcing ($Q_{eff}$), the outer regions become
marginally unstable to form spiral arms. Star formation occurs in these arms with higher local densities than the surrounding average.
Generally spirals grow at higher $Q$ than the radial stability threshold, so the spirals are the primary instability,
from which star formation follows at the locally increased densities in the arms.

We have traced smooth exponential stellar disks over 3-orders of magnitude and 4-6 disk scale lengths,
in spite of a highly variable gravitational instability parameter. Thus, gravitational instability thresholds
do not seem relevant to the stellar disk.
One possibility for creating an exponential disk is that
the star formation activity has an exponential profile and this forces the stellar disk to build up such a profile too.
Another possibility is that the stellar disk is continuously adjusted to an exponential shape regardless of the star formation profile.
The latter could be due to global dynamical processes that scatter the stars.
However, the known scattering processes only operate in spiral systems and cannot explain the same situation of smooth exponential
disks observed in \dirr\ galaxies.
Perhaps clumps in \dirr\ galaxies play the role of scattering centers to make the disks exponential (Hunter \et\ 2011).

\acknowledgments
We are grateful to J. M. van der Hulst for help in retrieving the \HI\ data of UGC 2885 from the WHISP archives.
The WHISP observation were carried out with the Westerbork Synthesis Radio Telescope, which is operated by
the Netherlands Foundation for Research in Astronomy (ASTRON) with financial support from the
Netherlands Foundation for Scientific Research (NWO). The WHISP project was carried out at the
Kapteyn Astronomical Institute by J.\ Kamphuis, D.\ Sijbring and Y.\ Tang under the supervision of
T.\ S.\ van Albada, J.\ M.\ van der Hulst and R.\ Sancisi.
Support for this work was provided to DAH by the Mt.\ Cuba Astronomical Foundation and by grant AST-0707563 from
the National Science Foundation. BGE received support from National Science Foundation grant AST-0707426.
AA and TW are grateful for participation in the Northern Arizona University Research
Experiences for Undergraduates program in the summers of 2012 and 2010.
This program is run by Dr.\ Kathy Eastwood and funded by the National Science Foundation through grant AST-1004107.
This research has made use of NED which is operated by the
Jet Propulsion Laboratory, California Institute of Technology, under contract with the National Aeronautics and Space Administration.

Facilities: \facility{Hall 1.1 m}, \facility{KPNO}, \facility{WSRT}



\clearpage

\clearpage

\clearpage

\clearpage

\clearpage

\clearpage

\clearpage

\clearpage

\clearpage

\clearpage

\clearpage

\clearpage

\clearpage

\clearpage

\clearpage

\clearpage

\begin{deluxetable}{lcccc}
\tabletypesize{\scriptsize}
\tablecaption{Galaxy Characteristics
\label{tab-gal}}
\tablewidth{0pt}
\tablehead{
\colhead{Parameter} & \colhead{NGC 801} & \colhead{UGC 2885}
}
\startdata
D (Mpc)\tablenotemark{a}                                                             &  79.4     & 79.1 \\
Center (h m s, deg arcmin arcsec; J2000)\tablenotemark{b} &  2:03:45.4 38:15:32 & 3:53:02.5 35:35:22  \\
PA (deg)\tablenotemark{b}                                                            & $-$28  &  47.5 \\
$b/a$\tablenotemark{b}                                                                  & 0.24     &  0.33 \\
Incl (deg)\tablenotemark{b}                                                           &  82       &  74     \\
E(B$-$V)$_f$\tablenotemark{c}                                                    & 0.042   & 0.176 \\
E(B$-$V)$_i^{HII}$\tablenotemark{c}                                          & 0.18     &  0.15 \\
M$_V$\tablenotemark{d}                                                               & $-$21.99$\pm$0.005  & $-$22.71$\pm$0.007 \\
log $M_{HI}$ (M\solar)                                                                   & 10.48 & 10.62 \\
$R_{1/2}$ (kpc)\tablenotemark{e}                                               & 15.4 & 22.2 \\
$R_{25}$ (kpc)\tablenotemark{f}                                                 & 17.3  & 22.2 \\
$R_D$ (kpc)\tablenotemark{g}                                                      & 14.14$\pm$0.52 & 12.05$\pm$0.41 \\
$\mu_V^0$ (mag of 1  arcsec$^2$)\tablenotemark{h}             & 23.09$\pm$0.14 & 21.26$\pm$0.15 \\
$\log L_{H\alpha}$ (ergs s$^{-1}$)\tablenotemark{i}              & 41.6  &  41.5 \\
log SFR$_{H\alpha}$ (M\solar yr$^{-1}$)\tablenotemark{j}                            & 0.50   & 0.41 \\
\enddata
\tablenotetext{a}{Determined from V$_{GSR}$, taken from NED and a Hubble constant of 75 km s$^{-1}$ Mpc$^{-1}$.}
\tablenotetext{b}{Determined from the $V$-band image. The inclination assumes an intrinsic $b/a$ of 0.2.}
\tablenotetext{c}{Foreground reddening E(B$-$V)$_f$ from Schlafly \& Finkbeiner (2011). Internal reddening
E(B$-$V)$_i^{HII}$ from Prescott \et\ (2007) for star-forming regions. For the field star E(B$-$V)$_i^*$, we multiply E(B$-$V)$_i^{HII}$
by 0.44 (Calzetti 1997).
The value of E(B$-$V)$_i^{HII}$ given here is evaluated at the $V$-band half-light radius $R_{1/2}$, determined
before the reddening corrections were applied.}
\tablenotetext{d}{Corrected for foreground extinction using A$_{V,f}$/E(B$-$V)$_f$=3.1 (Cardelli \et\ 1989) and the internal extinction
determined at the half light radius using A$_{V,i}$/E(B$-$V)$_i^*$=4.05 (Calzetti \et\ 1994, 2000).}
\tablenotetext{e}{Half-light radius of galaxy in the $V$-band not corrected for extinction.}
\tablenotetext{f}{Radius of galaxy at $B$-band isophote of 25 mag of 1 arcsec$^{2}$ before the extinction correction was applied.}
\tablenotetext{g}{Disk scale length determined from the fit to the outer disk $V$-band surface photometry, corrected for extinction.
$R_D$ for NGC 801 is for the outer component of the surface brightness profile.}
\tablenotetext{h}{Central surface brightness of the stellar disk from the fit to the outer disk $V$-band surface photometry.}
\tablenotetext{i}{Integrated \ha\ luminosity, corrected for extinction using the value appropriate to $R_{1/2}$. This value
has also been corrected for [NII] emission in the filter bandpass (see \S 3.3.1).}
\tablenotetext{j}{Integrated SFR determined from $L_{H\alpha}$ using the formula of Kennicutt (1998) (see \S 3.3.3).}
\end{deluxetable}


\begin{thebibliography}{}

\bibitem[Asplund et al.(2009)]{asplund09} Asplund, M., Grevesse, N., Sauval, A. J., \& Scott, P. 2009, ARA\&A, 47, 481
\bibitem[Barker et al.(2012)]{barker12} Barker, M. K., Ferguson, A. M. N., Irwin, M. J., Arimoto, N., \& Jablonka, P.
2012, MNRAS, 419, 1489
\bibitem[Barnes et al.(2012)]{barnes12} Barnes, K.\ L., van Zee, L., C\^ot\'e, S., \& Schade, D.\ 2012, ApJ, 757, 64
\bibitem[Beckman et al.(2000)]{beckman00} Beckman, J. E., Rozas, M., Zurita, A., Watson, R. A., \& Knapen, J. H. 2000, AJ, 119, 2728
\bibitem[Bell \& de Jong(2000)]{bell2000} Bell, E. F., \& de Jong, R. S. 2000, MNRAS, 312, 497
\bibitem[Bertin et al.(1989)]{bertin89} Bertin, G., Lin, C. C., Lowe, S. A., \& Thurstans, R. P. 1989, ApJ, 338, 104
\bibitem[Bigiel et al.(2008)]{bigiel08} Bigiel, F., Leroy, A., Walter, F., Brinks, E., de Blok, W. J. G., et al. 2008, AJ, 136, 2846
\bibitem[Bigiel et al.(2010)]{bigiel10} Bigiel, F., Leroy, A., Walter, F., Blitz, L., Brinks, E., de Blok, W. J. G., \& Madore, B. 2010, AJ, 140, 1194
\bibitem[Bland-Hawthorn et al.(2005)]{bland05} Bland-Hawthorn, J., Vlaji\'{c}, M., Freeman, K. C., \& Draine, B. T. 2005, ApJ, 629, 239
\bibitem[Bossier et al.(2007)]{bossier07} Bossier, S., Gil de Paz, A., Boselli, A., Madore, B. F., Buat, V., et al.\ 2007, ApJS, 173, 524
\bibitem[Brinchman et al.(2004)]{brinchman04}  Brinchmann, J., Charlot, S., White, S. D. M., et al. 2004, MNRAS, 351, 1151
\bibitem[Burstein et al.(1982)]{burstein82} Burstein, D., Rubin, V. C., Thonnard, N., \& Ford, W. K., Jr. 1982, ApJ, 253, 70
\bibitem[Calzetti(1997)]{calzetti97} Calzetti, D.\ 1997, in AIP Conf. Proc. 408, The Ultraviolet Universe at Low and High Redshift : Probing the Progress of Galaxy Evolution, eds. W. H. Waller, M. N. Fanelli, J. E. Hollis, \& A. C. Danks (Woodbury:AIP), 403
\bibitem[Calzetti et al.(2000)]{calzetti00} Calzetti, D., Armus, L., Bohlin, R. C., Kinney, A. L., Koornneef, J., \& Storchi-Bergmann, T. 2000, ApJ, 533, 682
\bibitem[Calzetti et al.(1994)]{calzetti94} Calzetti, D., Kinney, A. L., \& Storchi-Bergmann, T. 1994, ApJ, 429, 582
\bibitem[Cardelli et al.(1989)]{cardelli89} Cardelli, J.\ A., Clayton, G.\ C., \& Mathis, J.\ S.\ 1989, ApJ, 345, 245
\bibitem[Carraro et al.(2010)]{carraro10} Carraro, G., V\'azquez, R. A., Costa, E., Perren, G., \& Moitinho, A. 2010, ApJ, 718, 683
\bibitem[Chang et al.(2012)]{chang12} Chang, R. X., Shen, S. Y., \& Hou, J. L. 2012, ApJL, 753, id L10
\bibitem[Chiappini et al.(1997)]{chiappini97} Chiappini, C., Matteucci, F., \& Gratton, R. 1997, ApJ, 477, 765
\bibitem[Christlein et al.(2010)]{christlein10} Christlein, D., Zaritsky, D., \& Bland-Hawthorn, J. 2010, MNRAS, 405, 2549
\bibitem[de Vaucouleurs \& de Vaucouleurs (1972)]{deV72} de Vaucouleurs, G., \& de Vaucouleurs, A. 1972, Memoirs of the Royal Astronomical Society, Vol 77, p 1
\bibitem[de Vaucouleurs et al.(1991)]{rc3} de Vaucouleurs, G., de Vaucouleurs, A., Corwin, H., Buta, R., Paturel, G., \& Fouqu\'e, P.\ 1991, Third Reference Catalogue of Bright Galaxies (New York, Springer-Verlag) (RC3)
\bibitem[Elmegreen(2011)]{elmegreen11} Elmegreen, B. G. 2011, ApJ, 737, 10
\bibitem[Espada et al.(2011)]{espada11} Espada, D., Mu\~noz-Mateos, J. C., Gil de Paz, A., et al. 2011, ApJ, 736, 20
\bibitem[Eufrasio et al.(2013)]{eufrasio13} Eufrasio, R. T., de Mello, D. F., Urrutia-Viscarra, F., de Oliveira, C. M., \& Dwek, E. 2013, in Proceedings of the International Astronomical Union, Volume 8, Symposium S292, ``Molecular Gas, Dust, and Star Formation in Galaxies,'' p 328
\bibitem[Ferguson \& Johnson(2001)]{ferguson01} Ferguson, A. M. N., \& Johnson, R. A. 2001, ApJ, 559, L13
\bibitem[Ferguson et al.(1998)]{ferguson98} Ferguson, A. M. N., Wyse, R. F. G., Gallagher, J. S., \& Hunter, D. A. 1998, ApJ, 506, 19
\bibitem[Grossi et al.(2011)]{grossi11} Grossi, M., Hwang, N., Corbelli, E., Giovanardi, C., Okamoto, S., \& Arimoto, N. 2011, A\&A, 533, A91
\bibitem[Henry \& Worthey(1999]{henry99} Henry, R. B. C., \& Worthey, G. 1999, PASP, 111, 919
\bibitem[Herbert-Fort et al.(2012)]{herbert12} Herbert-Fort, S., Zaritsky, D., Moustakas, J., Di Paola, A., Pogge, R. W., \& Ragazzoni, R.
2012, ApJ, 754, id 110
\bibitem[Herrmann et al.(2013)]{herrmann13} Herrmann, K. A., Hunter, D. A., \& Elmegreen, B. G. 2013, AJ, submitted
\bibitem[Holwerda et al.(2005)]{holwerda05} Holwerda, B. W., Gonz\'alez, R. A., van der Kruit, P. C., \& Allen, R. J. 2005, A\&A, 444, 109
\bibitem[Huizinga \& van Albada(1992)]{huizinga92} Huizinga, J. E. \& van Albada, T. S., 1992, MNRAS, 254, 677
\bibitem[Hunter(1982)]{hunter82} Hunter, D. A. 1982, ApJ, 260, 81
\bibitem[Hunter et al.(2010)]{hunter10} Hunter, D. A., Elmegreen, B. G., \& Ludka, B. C. 2010, AJ, 139, 447
\bibitem[Hunter et al.(2011)]{hunter11} Hunter, D. A., Elmegreen, B. G., Oh, S.-H., et al.\ 2011, AJ, 142, 121
\bibitem[Hunter \& Gallagher(1986)]{hunter86}  Hunter, D. A., \& Gallagher, J. S., III 1986, PASP, 98, 5
\bibitem[James et al.(2005)]{james05} James, P. A., Shane, N. S., Knapen, J. H., Etherton, J., Percival, S. M. 2005, A\&A, 429, 851
\bibitem[Jog \& Solomon(1984)]{jog84} Jog, C. J., \& Solomon, P. M. 1984, ApJ, 276, 114
\bibitem[Julian \& Toomre(1966)]{julian66} Julian, W. H., \& Toomre, A. 1966, ApJ, 146, 810
\bibitem[Kamphuis \& Sancisi(1993)]{kamphuis93} Kamphuis, J., \& Sancisi, R. 1993, A\&A, 273, L31
\bibitem[Kennicutt(1984)]{Kennicutt84} Kennicutt, R. C., Jr. 1984, ApJ, 287, 116
\bibitem[Kennicutt(1998)]{Kennicutt98} Kennicutt, R. C., Jr. 1998, ARAA, 36, 189
\bibitem[Kennicutt(1989)]{Kennicutt89} Kennicutt, R. C., Jr. 1989, ApJ, 344, 685
\bibitem[Kennicutt et al.(2007)]{kennicutt07} Kennicutt, R. C., Jr., Calzetti, D., Walter, F. et al. 2007, ApJ, 671, 333
\bibitem[Landolt(1992)]{landolt92} Landolt, A.\ U.\ 1992, AJ, 104, 340
\bibitem[Larson(1976)]{larson76} Larson, R. B. 1976, MNRAS, 176, 31
\bibitem[Lau \& Bertin(1978)]{lau78} Lau, Y. Y., \& Bertin, G. 1978, ApJ, 226, 508
\bibitem[Lee et al.(2011)]{lee11} Lee, J. H., Hwang, N., \& Lee, M. G. 2011, ApJ, 735, 75
\bibitem[Leitherer et al.(1999)]{leitherer99} Leitherer, C., Schaerer, D., Goldader, J. D., Gonz\'alez Delgado, R. M.,
Robert, C., et al.\ 1999, ApJS, 123, 3
\bibitem[Leroy et al.(2008)]{leroy08} Leroy, A. K., Walter, F., Brinks, E., et al. 2008, ApJ, 136, 2782
\bibitem[Mestel et al.(1963)]{mestel63} Mestel, L. 1963, MNRAS, 126, 553
\bibitem[Mo et al.(1998)]{mo98} Mo, H. J., Mao, S., \& White, S. D. M. 1998, MNRAS, 295, 319
\bibitem[Nab \& Ostriker(2006)]{nab06} Naab, T. \& Ostriker, J. P. 2006, MNRAS, 366, 899
\bibitem[Osterbrock(1989)]{osterbrock89} Osterbrock, D. E. 1989, Astrophysics of Gaseous Nebulae and Active Galactic Nuclei
(Mill Valley, California:University Science Books)
\bibitem[Ostriker et al.(2010)]{ostriker10} Ostriker, E. C., McKee, C. F., \& Leroy, A. K. 2010, ApJ,  721, 975
\bibitem[Patterson et al.(2012)]{patterson12} Patterson, M. T., Walterbos, R. A. M., Kennicutt, R. C., Chiappini, C., \&Thilker, D. A. 2012, MNRAS, 422, 401
\bibitem[Pellegrini et al.(2012)]{pellegrini12} Pellegrini, E. W., Oey, M. S., Winkler, P. F., et al. 2012, ApJ, 755, 40
\bibitem[Pilyugin et al.(2007)]{pilyugin07} Pilyugin, L.\ S., Thuan, T.\ X., \& V'lchez, J.\ M. 2007, MNRAS, 375, 353
\bibitem[Prescott et al.(2007)]{prescott07} Prescott, M. K. M., Kennicutt, R. C., Jr., Bendo, G. J., \et\. 2007, ApJ, 668, 182
\bibitem[Quirk(1972)]{quirk72} Quirk, W. J. 1972, ApJL, 176, L9
\bibitem[Radburn-Smith et al.(2012)]{radburn12}  Radburn-Smith, D. J., Ro\u{s}kar, R., Debattista, V. P., Dalcanton, J. J., Streich, D., de 
Jong, R.S., Vlajic, M., Holwerda, B.W., Purcell, C.W., Dolphin, A.E., \& Zucker, Daniel B.  2012, ApJ, 753, 138 
\bibitem[Rela\~no et al.(2012)]{relano12} Rela\~no, M., Kennicutt, R. C., Jr., Eldridge, J. J., Lee, J. C., \& Verley, S.
2012, MNRAS, 423, 2933
\bibitem[Roediger et a.(2012)]{roediger12} Roediger, J. C., Courteau, S., S\'anchez-Bl\'azquez, P., \& McDonald, M. 2012, ApJ, 758, 41
\bibitem[Roelfsema \& Allen(1985)]{roelf85} Roelfsema, P. R., \& Allen, R. J. 1985, A\&A, 146, 213
\bibitem[Romeo \& Wiegert(2011)]{romeo11} Romeo, A. B., \& Wiegert, J. 2011, MNRAS, 416, 1191
\bibitem[Ro\"okar et al.(2008)]{rookar08} Ro\"okar, R., Debattista, V. P., Stinson, G. S., Quinn, T. R., Kaufmann, T., \& Wadsley, J. 2008, ApJ, 675, L65
\bibitem[Ro\u{s}kar et al.(2008)]{roskar08} Ro\u{s}kar, R., Debattista, V. P., Quinn, T. R., Stinson, G. S., \& Wadsley, J. 2008, ApJ, 684, L79
\bibitem[Rubin et al.(1980)]{rubin80} Rubin, V. C., Ford, W. K. Jr., \& Thonnard, N. 1980, ApJ, 238, 471
\bibitem[Ryder \& Dopita(1994)]{ryder94} Ryder, S. D., \& Dopita, M. A. 1994, ApJ, 430, 142
\bibitem[Saha et al.(2010)]{saha10} Saha, A., et al. 2010, AJ, 140, 1719
\bibitem[Salpeter(1955)]{salpeter55} Salpeter, E. E. 1955, ApJ, 121, 161
\bibitem[Sault et al.(1995)]{sault95} Sault, R. J., Teuben, P. J., \& Wright, M. C. H. 1995, in ASP Conference Series Vol 77, Astronomical Data Analysis Software and Systems IV, eds. R. A. Shaw, H. E. Payne, \& J. J. E. Hayes (San Francisco:Astronomical Society of the Pacific), p 433
\bibitem[Schlafly \& Finkbeiner(2011))]{schlafly11} Schlafly, E. F. \& Finkbeiner, D. P. 2011, ApJ, 737, 103
\bibitem[Schlegel et al.(1998)]{schlegel98} Schlegel, D. J., Finkbeiner, D. P., \& Davis, M. 1998, ApJ, 500, 525
\bibitem[Sellwood \& Binney(2002)]{sellwood02} Sellwood, J. A., \& Binney, J. J. 2002, MNRAS, 336, 785
\bibitem[Tamburro et al.(2009)]{tamburro09} Tamburro, D., Rix, H.-W., Leroy, A. K., et al. 2009, AJ, 137, 4424
\bibitem[Thilker et al.(2005)]{thilker05} Thilker, D. A., Bianchi, L., Boissier, S., Gil de Paz, A., Madore, B. F., et al.\ 2005, ApJ, 619, L79
\bibitem[Thilker et al.(2007)]{thilker07} Thilker, D. A., Bianchi, L., Meurer, G., Gil de Paz, A., Boissier, S., et al.\ 2007, ApJS, 173, 538
\bibitem[Toomre(1964)]{toomre64} Toomre, A.\ 1964, ApJ, 139, 1217
\bibitem[van der Hulst et al.(2001)]{hulst01} van der Hulst, J. M., van Albada, T. S., \& Sancisi, R. in ASP Conference Proceedings Vol 240, Gas and Galaxy Evolution, eds. J. E. Hibbard, M. Rupen, \& J. H. van Gorkom (San Francisco:Astronomical Society of the Pacific), p 451
\bibitem[van der Kruit(1987)]{kruit87} van der Kruit, P. C. 1987, A\&A, 173, 59
\bibitem[van der Kruit \& Freeman(2011)]{kruit11} van der Kruit, P. C., \& Freeman, K. 2011, ARA\&A, 49, 301
\bibitem[Vlaji\'{c} et al.(2009)]{vlajic09} Vlaji\'{c}, M., Bland-Hawthorn, J., \& Freeman, K. C. 2009, ApJ, 697, id 361
\bibitem[Vlaji\'{c} et al.(2011)]{vlajic11} Vlaji\'{c}, M., Bland-Hawthorn, J., \& Freeman, K. C. 2011, ApJ, 732, id 7
\bibitem[Vogelaar \& Terlouw(2001)]{gipsy01} Vogelaar, M. G. R., \& Terlouw, J. P. 2001, in Astronomical
Data Analysis Software Systems I, eds D. M. Worall, C. Biemesderfer, and J. Barnes, ASP Conference Serices No.\ 25, p 131
\bibitem[Wang \& Heckman(1996)]{wang96} Wang, B., \& Heckman, T. M., 1996, ApJ, 457, 645
\bibitem[Wang et al.(2013)]{wang13} Wang, J., Kauffmann, G., J\'{o}zsa, G. I. G., et al.\ 2013, MNRAS, submitted
\bibitem[Yoachim et al.(2012)]{yoachim12} Yoachim, P., Ro\v{s}kar, \& Debattista, V. P. 2012, ApJ, 752, id 97
\bibitem[Zasov \& Simakov(1988)]{zasov88} Zasov, A. V., \& Simakov, S. G. 1988, Astrophysics, 29, 518
\end{thebibliography}
\end{document}